\documentclass[%
 reprint,
superscriptaddress,
preprintnumbers,
 amsmath,amssymb,
aps,
pra,
]{revtex4-2}

\usepackage{CJK}
\usepackage{enumitem}

\usepackage{amsmath}
\usepackage{physics}
\usepackage{braket}
\usepackage{graphicx}
\usepackage{dcolumn}
\usepackage{bm}

\usepackage{graphicx}
\usepackage{overpic}

\usepackage{xcolor} 
\usepackage[colorlinks=true]{hyperref} 

\newcounter{algorithm}

\def\be{\begin{equation}}
\def\ee{\end{equation}}
\def\ba{\begin{eqnarray}}
\def\ea{\end{eqnarray}}
\def\bi{\begin{itemize}}
\def\ei{\end{itemize}}

\begin{document}


\title{
Quantum Path Finder: Navigating Constraint Space via Non-Abelian Gauge Dynamics
 }


\begin{CJK}{UTF8}{gbsn}
\author{Chunyin Chen (陈椿尹)}
\affiliation{International Center for Quantum Materials, School of Physics, Peking University, Beijing 100871, China}

\author{Biao Wu (吴飙)}
\email{wubiao@pku.edu.cn}
\affiliation{International Center for Quantum Materials, School of Physics, Peking University, Beijing 100871, China}
\affiliation{Wilczek Quantum Center, Shanghai Institute for Advanced Studies, University of Science and Technology of China, Shanghai 201315, China}
\affiliation{Hefei National Laboratory, Hefei 230088, China}

\date{\today}

\begin{abstract}
Quantum algorithms are commonly designed within the circuit paradigm or through adiabatic evolution. Here we introduce  \emph{Quantum Path Finder} (QPF), a distinct algorithm-design paradigm based on non-Abelian gauge dynamics in a constraint-protected degenerate manifold. Each Boolean variable is promoted to a qutrit whose additional state represents an unassigned variable. The problem Hamiltonian then defines a connected manifold of feasible partial assignments, and time-dependent local qutrit rotations generate coherent transport on this manifold. Unlike conventional adiabatic quantum computation, QPF does not require the state to follow an instantaneous eigenstate along a prescribed interpolation: adiabaticity with respect to the constraint gap protects the feasible manifold, while adiabatic following within that manifold is optional. The algorithmic task is therefore to design paths and interference in constraint space rather than only to engineer a sequence of gates or a ground-state interpolation. We illustrate the framework with three applications. For unstructured search, a two-level analysis and numerical evolution recover Grover-type scaling under explicitly stated assumptions. For complete $3$-XORSAT, the accessible gauge sector has a size-independent gap, in contrast to the exponentially closing gap of a conventional transverse-field adiabatic path. For satisfiable $3$-SAT instances, finite-size benchmarks demonstrate high satisfied-clause fractions at short classical comparison budgets, without claiming an equal-cost quantum advantage. We also outline a superconducting-qutrit implementation and discuss the conceptual relationship between QPF and nondeterministic Turing machine.
\end{abstract}

\maketitle
\end{CJK}

\section{Introduction}
\label{sec:introduction}
Quantum circuit algorithms and quantum adiabatic algorithms are two dominant paradigms for quantum-algorithm design. In the circuit paradigm, a computation is decomposed into a sequence of unitary gates acting on a quantum register, followed by measurement; the algorithm is specified by the choice, ordering, and composition of these gates~\cite{Chuang}. In the adiabatic paradigm, the system is initialized in an easily prepared ground state and then evolves sufficiently slowly under a time-dependent Hamiltonian, so that it remains close to the instantaneous ground state while the Hamiltonian is transformed into one whose ground state encodes the solution~\cite{KadowakiNishimori1998,Farhi2000Adiabatic,Farhi2001,AlbashLidar2018}. These paradigms offer complementary perspectives on algorithm design: discrete gate construction in the circuit model and continuous ground-state-path engineering in adiabatic computation.

Here we develop the \emph{Quantum Path Finder} (QPF) as a distinct paradigm for quantum-algorithm design. Its central object is not a circuit decomposition or a single adiabatic eigenstate path, but coherent transport through a degenerate manifold of computational configurations. A problem is first lifted to an enlarged Hilbert space containing partial as well as complete assignments. Constraints select a degenerate zero-energy manifold, and controlled local rotations generate a non-Abelian gauge potential within that manifold~\cite{WilczekZee1984,ZanardiRasetti1999Holonomic}. The resulting dynamics explores paths through constraint space, with interference among paths forming an essential part of the algorithm. This construction generalizes non-Abelian adiabatic mixing for independent-set problems to arbitrary Boolean constraints and to manifolds containing all feasible partial assignments~\cite{WuYuWilczek2020,YuWilczekWu2021,Zhao2025MIS,Perlin2025QCHOP}.

The distinction from conventional adiabatic quantum computation is crucial. In a standard adiabatic algorithm, the computational state must remain close to an instantaneous ground state, or another selected eigenstate, while the Hamiltonian interpolates between an initial driver and the problem Hamiltonian~\cite{KadowakiNishimori1998,Farhi2000Adiabatic,Farhi2001,AlbashLidar2018}. The minimum gap along this interpolation is therefore often the principal bottleneck. QPF instead uses an isospectral family $U(t)H_{\rm P}U^\dagger(t)$ to move an entire degenerate feasible manifold. Slowness relative to the constraint gap suppresses leakage out of this manifold, but the state need not follow an instantaneous eigenstate of the effective Hamiltonian within it. A second adiabatic condition can be imposed on the internal gauge dynamics, yet it is not required by the general algorithm. Consequently, QPF permits both adiabatic and nonadiabatic transport inside the feasible manifold, and its design freedom resides in the connectivity, on-site potentials, and hopping amplitudes of an effective graph rather than solely in the gap of a prescribed eigenstate path.

Partial assignments are central to this construction. We encode each Boolean variable as a qutrit with an auxiliary unassigned state $\ket u$, in addition to the Boolean states $\ket0$ and $\ket1$. A constraint is penalized only 
when assigned variables  violate that constraint. The zero-energy manifold consequently contains every feasible partial assignment. Because unassigning a variable cannot create a violation, all such configurations are connected by local assignment and unassignment moves. Assignment depth provides a natural coordinate on this connected graph, from the trivial root $\ket u^{\otimes n}$ to complete solutions when they exist. Local qutrit controls then generate both depth-changing transport and, if desired, same-depth mixing. The algorithm can therefore be viewed as navigating a constraint-organized configuration space rather than searching only among complete assignments.

We establish the general formalism and examine it in three representative settings. For unstructured search, an explicit boundary-coupling analysis yields Grover-type gap and runtime scaling under isolated-crossing and finite-overlap assumptions, supported by numerical time evolution. This demonstrates that QPF can be as powerful 
as quantum circuit algorithms and quantum adiabatic algorithms. 

For satisfiable complete $3$-XORSAT, we prove a uniform, size-independent lower bound on the gap in the accessible permutation-invariant gauge sector; this contrasts with the exponentially small gap of the conventional transverse-field adiabatic path considered here. For satisfiable $3$-SAT instances, finite-size simulations show high mean satisfied-clause fractions and high probabilities of satisfying at least $95\%$ of the clauses relative to several classical baselines at a short clause-inspection budget. These benchmarks are not equal-resource comparisons and do not by themselves establish asymptotic quantum advantage. We also present a fixed-basis spin-$1$ Hamiltonian and outline how its local drives and interactions could be realized with superconducting qutrits.

The remainder of the paper is organized as follows. Section~\ref{sec:qutrit-encoding} defines the qutrit encoding and the geometry of the feasible manifold. Section~\ref{sec:gauge} derives the non-Abelian gauge dynamics, specifies the QPF protocol and its controls, and discusses a possible physical implementation. Section~\ref{sec:applications} presents the applications to unstructured search, $k$-XORSAT, and $k$-SAT. Section~\ref{sec:discussion} discusses the algorithmic interpretation, including the analogy with and differences from nondeterministic Turing machines, together with limitations and open directions. Section~\ref{sec:conclusion} concludes. Technical derivations, additional constraint encodings, and details of the classical benchmarks are given in the appendices.

\begin{figure*}[t]
\centering
\includegraphics[width=0.90\textwidth]{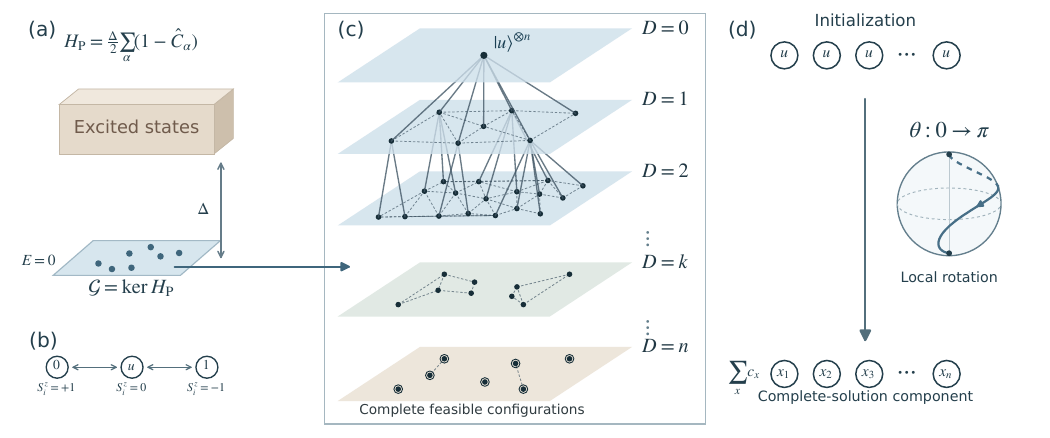}
\caption{
Qutrit encoding and non-Abelian mixing on the feasible partial-assignment manifold. (a) The positive-semidefinite penalty Hamiltonian $H_{\rm P}$ has a degenerate zero-energy manifold $\mathcal G$; constraint-violating configurations have energies of at least $\Delta$. (b) Local qutrit encoding: $|0\rangle$, $|u\rangle$, and $|1\rangle$ correspond to $S^z=+1,0,-1$, respectively. Arrows indicate the assignment and unassignment transitions $|u\rangle\leftrightarrow|0\rangle$ and $|u\rangle\leftrightarrow|1\rangle$. (c) Feasible configurations form a connected graph under these single-site moves. The graph is layered by assignment depth $D$, from the root $|u\rangle^{\otimes n}$ to the maximal feasible depth $D_{\max}$. The illustrated case $D_{\max}=n$ contains complete solutions; for an unsatisfiable instance, $D_{\max}<n$. (d) Time-dependent local qutrit rotations induce non-Abelian gauge dynamics within the feasible manifold, with controls designed to transport amplitude toward larger assignment depth.
}
\label{fig:feasible_layers}
\end{figure*}

\section{Qutrit Encoding and Feasible Subspace}
\label{sec:qutrit-encoding}

We begin by defining the qutrit representation and the feasible partial-assignment manifold illustrated in Fig.~\ref{fig:feasible_layers}.

\subsection{Qutrit encoding}

Consider a computational problem defined on $n$ Boolean variables $x_i\in\{0,1\}$, with target configurations specified by a problem Hamiltonian. To represent partial as well as complete configurations, we enlarge the local Hilbert space of each variable from a qubit to a qutrit,
\begin{equation}
\label{eq:qutrit-spin1-encoding}
\ket{u}_i\equiv\ket{S^z_i=0},\;
\ket{0}_i\equiv\ket{S^z_i=+1},\;
\ket{1}_i\equiv\ket{S^z_i=-1},
\end{equation}
where $\ket{0}_i$ and $\ket{1}_i$ encode the two Boolean values, while $\ket{u}_i$ denotes an auxiliary unassigned state, indicating that the value of $x_i$ has not yet been specified. Here $S_i^{x,y,z}$ are the standard spin-$1$ operators with $\hbar=1$. In the local basis $\{\ket{u}_i,\ket{0}_i,\ket{1}_i\}$, we use the convention
\begin{align}
S_i^z
&=
\ket{0}_i\bra{0}
-
\ket{1}_i\bra{1},
\\
S_i^+
&=
\sqrt{2}
\left(
\ket{0}_i\bra{u}
+
\ket{u}_i\bra{1}
\right),
\qquad
S_i^-=(S_i^+)^\dagger,
\end{align}
where
\begin{equation}
S_i^x=\frac{S_i^++S_i^-}{2},
\qquad
S_i^y=\frac{S_i^+-S_i^-}{2i}.
\end{equation}

The local qutrit basis defines the Hilbert space $\mathcal H_i=\mathrm{span}\{\ket{u}_i,\ket{0}_i,\ket{1}_i\}$, and the full Hilbert space is $\mathcal H=\bigotimes_{i=1}^n\mathcal H_i$. A computational-basis state is written as
\begin{equation}
\ket{\mu}
=
\ket{\mu_1,\ldots,\mu_n},
\qquad
\mu_i\in\{u,0,1\},
\end{equation}
where each $\mu_i$ specifies whether the $i$th variable is unassigned or takes one of the two Boolean values. Thus, $\ket{\mu}$ represents a partial configuration of the original Boolean variables. States with no unassigned variables correspond to the original Boolean configuration space.

\subsection{Problem Hamiltonian and feasible subspace}

We specify the problem by Boolean constraints $\{C_\alpha\}$, where $C_\alpha:\{0,1\}^{|\partial\alpha|}\rightarrow\{0,1\}$ acts on a nonempty subset $\partial\alpha\subseteq\{1,\ldots,n\}$ and $C_\alpha=1$ denotes satisfaction. The support $\partial\alpha$ may contain only a few variables, as in $k$-SAT, or all $n$ variables, as for the oracle in the unstructured-search example of Sec.~\ref{sec:unstructured-search}. We lift each classical constraint to a qutrit operator $\hat C_\alpha$ satisfying
\begin{equation}
\hat C_\alpha\ket{\mu}
=
\begin{cases}
+\ket{\mu}, & \text{if $\mu$ does not violate $C_\alpha$},\\
-\ket{\mu}, & \text{if $\mu$ violates $C_\alpha$}.
\end{cases}
\label{eq:constraint-operator}
\end{equation}
A constraint is violated only when every variable in its support is assigned and their Boolean values violate the constraint; otherwise, including when any participating variable is unassigned, it is nonviolated. The problem Hamiltonian is
\begin{equation}
H_{\rm P}
=
\frac{\Delta}{2}
\sum_\alpha
\left(
1-\hat C_\alpha
\right),
\qquad
\Delta>0.
\label{eq:problem-hamiltonian}
\end{equation}

We use a uniform energy scale $\Delta$ for simplicity, although constraint-dependent weights can be introduced directly. Each term in $H_{\rm P}$ is positive semidefinite and vanishes on partial configurations that do not violate the corresponding constraint. The zero-energy manifold is therefore
\begin{equation}
\mathcal{G}
=
\ker H_{\rm P}
=
\mathrm{span}
\left\{
\ket{\mu}:
\hat C_\alpha\ket{\mu}
=
\ket{\mu}
\;\text{for all}\;\alpha
\right\}.
\label{eq:feasible-manifold}
\end{equation}
Thus $\mathcal{G}$ is spanned by all feasible partial configurations, as illustrated in Fig.~\ref{fig:feasible_layers}(a). Every state outside $\mathcal{G}$ violates at least one constraint, so the feasible manifold is separated from the constraint-violating sector by an energy gap of at least $\Delta$. Feasibility, however, does not guarantee that a partial assignment admits a satisfying completion.

Table~\ref{tab:boolean-qutrit-lifting} in Appendix~\ref{app:hamiltonian-instances} gives explicit qutrit violation projectors for the applications studied below and several additional constraints.

We call the local transitions $\ket{u}\leftrightarrow\ket{0}$ and $\ket{u}\leftrightarrow\ket{1}$, which change one variable at a time [Fig.~\ref{fig:feasible_layers}(b)], \emph{single-site assignment flips}. These moves define the feasible graph: its vertices are feasible computational-basis configurations, and its edges connect configurations related by one such flip~\cite{Nishimura2018} [Fig.~\ref{fig:feasible_layers}(c)].

Every feasible partial configuration is connected to the root state $\ket{u}^{\otimes n}$. Indeed, unassigning one variable cannot create a violation: a nonviolated constraint either remains fully assigned and nonviolated or becomes partially assigned and hence nonviolated by definition. Repeating this operation connects any feasible configuration to the root, proving that the feasible graph is connected.

We define $d_{\mathcal G}(\mu,\nu)$ as the graph distance between two feasible configurations $\mu$ and $\nu$, namely the length of the shortest path connecting them on the feasible graph. We further define the assignment depth
\begin{equation}
D(\mu)
=
\#\{i:\mu_i\neq u\},
\label{eq:assignment-depth}
\end{equation}
which counts the number of assigned variables. Since any feasible configuration $\mu$ can be connected to the root state in $D(\mu)$ single-site assignment flips, any two feasible configurations satisfy
\begin{equation}
d_{\mathcal G}(\mu,\nu)
\leq
D(\mu)+D(\nu).
\end{equation}
In particular, for two configurations in the same depth-$D$ sector,
\begin{equation}
d_{\mathcal G}(\mu,\nu)\leq 2D.
\end{equation}

The feasible manifold naturally inherits the assignment-depth decomposition
\begin{equation}
\mathcal{G}
=
\bigoplus_{D=0}^{n}
\mathcal{G}_D,
\label{eq:depth-decomposition}
\end{equation}
where $\mathcal{G}_D$ is spanned by all feasible partial configurations with depth $D$. We define the maximal feasible depth by
\begin{equation}
D_{\max}
=
\max\left\{
D:\mathcal G_D\neq\{0\}
\right\}.
\label{eq:max-feasible-depth}
\end{equation}
For a satisfiable instance, $D_{\max}=n$, whereas for an unsatisfiable instance $D_{\max}<n$.

The root state $\ket{u}^{\otimes n}$ belongs to $\mathcal{G}_0$. When $D_{\max}=n$, the sector $\mathcal{G}_n$ is spanned by complete configurations satisfying every constraint. The resulting layered structure is shown in Fig.~\ref{fig:feasible_layers}(c). This construction generalizes the independent-set manifold of the qubit maximum-independent-set problem~\cite{Pichler2018RydbergMIS,Ebadi2022RydbergMIS}: the qutrit feasible manifold unifies the root, all intermediate feasible partial configurations, and, for satisfiable instances, all complete solutions within one degenerate subspace.

\section{Non-Abelian Gauge Evolution}
\label{sec:gauge}

To generate non-Abelian dynamics within the degenerate feasible manifold $\mathcal G$, we introduce an isospectral family of Hamiltonians related by time-dependent local qutrit rotations,
\begin{equation}
H(t)
=
U(t)H_{\rm P}U^\dagger(t),
\label{eq:rotated-problem-hamiltonian}
\end{equation}
where $U(t)=\bigotimes_i U_i(t)$ and $U_i(t)$ acts only on the $i$th qutrit. Unitarity preserves the spectrum of $H_{\rm P}$, while the zero-energy manifold evolves according to
\begin{equation}
\mathcal{G}(t)
=U(t)\mathcal{G}.
\end{equation}

When $U(t)$ varies sufficiently slowly relative to the spectral gap between $\mathcal{G}(t)$ and the excited subspace, transitions out of the feasible manifold are suppressed~\cite{JansenRuskaiSeiler2007}. The residual dynamics within $\mathcal{G}(t)$ is governed by a non-Abelian gauge matrix~\cite{WilczekZee1984,WuYuWilczek2020,YuWilczekWu2021,Zhao2025MIS}. We now derive this matrix and show how it generates coherent transitions between feasible partial configurations.

\subsection{Non-Abelian gauge matrix}
\label{sec:nagmx}

In general, each $U_i$ may be an independent $SU(3)$ matrix with eight control parameters. Here we restrict all local rotations to the same three-parameter form. Specifically, in the computational basis $\{\ket{u}_i,\ket{0}_i,\ket{1}_i\}$, we choose $U_i(t)$ from the subgroup $U(2)\times U(1)\subset U(3)$:
\begin{equation}
U_i(t)
=
\begin{pmatrix}
-e^{i\phi}\cos\frac{\theta}{2}
&
\dfrac{e^{i\phi}}{\sqrt{2}}\sin\frac{\theta}{2}
&
\dfrac{e^{i\phi}}{\sqrt{2}}\sin\frac{\theta}{2}
\\[6pt]
\dfrac{1}{\sqrt{2}}\sin\frac{\theta}{2}
&
\dfrac{1}{2}\left(\cos\frac{\theta}{2}+e^{i\eta}\right)
&
\dfrac{1}{2}\left(\cos\frac{\theta}{2}-e^{i\eta}\right)
\\[6pt]
\dfrac{1}{\sqrt{2}}\sin\frac{\theta}{2}
&
\dfrac{1}{2}\left(\cos\frac{\theta}{2}-e^{i\eta}\right)
&
\dfrac{1}{2}\left(\cos\frac{\theta}{2}+e^{i\eta}\right)
\end{pmatrix},
\label{eq:local-qutrit-unitary}
\end{equation}
with $\theta=\theta(t)$, $\phi=\phi(t)$, and $\eta=\eta(t)$. Appendix~\ref{app:gauge-derivation} motivates this choice and gives the detailed derivation. The projected non-Abelian gauge matrix is
\begin{equation}
A(t)
=
iP_{\mathcal G}U^\dagger(t)\dot U(t)P_{\mathcal G},
\label{eq:projected-gauge}
\end{equation}
where $P_{\mathcal G}$ projects onto $\mathcal G$. Because $P_{\mathcal G}H_{\rm P}P_{\mathcal G}=0$, projection onto the feasible manifold yields the secondary Hamiltonian
\begin{equation}
    H_{\rm sec}(t)
    \equiv
    P_{\mathcal G}H_{1}(t)P_{\mathcal G}
    =
    -A(t),
    \label{eq:projected-effective-H}
\end{equation}
with $H_1(t)=-iU^\dagger(t)\dot U(t)$.
In the constraint-protected regime, this gauge Hamiltonian governs the effective dynamics within $\mathcal G$, which remains separated from the constraint-violating sector by an energy gap of at least $\Delta$.

Because $U(t)=\bigotimes_i U_i(t)$ is a product of local rotations, off-diagonal gauge matrix elements connect only feasible configurations that differ at a single site. The direct calculation in Appendix~\ref{app:gauge-derivation} gives, for a feasible partial configuration $\mu$, the diagonal matrix element
\begin{equation}
A_{\mu\mu}(t)
=
-(n-D(\mu))
\cos^2\frac{\theta}{2}\dot\phi
-
\frac{D(\mu)}{2}
\left(
\sin^2\frac{\theta}{2}\dot\phi
+
\dot\eta
\right).
\label{eq:gauge-diagonal}
\end{equation}
For two distinct feasible configurations $\mu$ and $\nu$, the off-diagonal matrix element is
\begingroup\small
\begin{equation}
A_{\mu\nu}(t)
=
\setlength{\arraycolsep}{3pt}
\begin{cases}
\dfrac{1}{\sqrt{2}}
\left(
\dfrac{1}{2}\sin\theta\,\dot\phi
-
\dfrac{i}{2}\dot\theta
\right),
&
\mu_i=u,\quad \nu_i\in\{0,1\},
\\[8pt]
\dfrac{1}{\sqrt{2}}
\left(
\dfrac{1}{2}\sin\theta\,\dot\phi
+
\dfrac{i}{2}\dot\theta
\right),
&
\nu_i=u,\quad \mu_i\in\{0,1\},
\\[8pt]
\dfrac{1}{2}
\left(
\dot\eta
-
\sin^2\frac{\theta}{2}\dot\phi
\right),
&
\{\mu_i,\nu_i\}=\{0,1\},
\\[6pt]
0,
&
\text{otherwise}.
\end{cases}
\label{eq:gauge-offdiagonal}
\end{equation}
\endgroup

These matrix elements determine the secondary Hamiltonian $H_{\rm sec}(t)=-A(t)$ that governs dynamics within the feasible manifold. The structure and control of $A(t)$ are central to QPF; we describe the general algorithm before examining specific control protocols.

\subsection{Quantum Path Finder Algorithm}
QPF is a general framework rather than an algorithm tailored to a single computational problem. Its five steps are as follows.
\begin{enumerate}
\item \textit{Encoding.}
Promote each Boolean variable to a qutrit with local basis $\{\ket{u},\ket{0},\ket{1}\}$. Encode the constraints in the problem Hamiltonian $H_{\rm P}=\frac{\Delta}{2}\sum_\alpha(1-\hat C_\alpha)$, whose zero-energy manifold $\mathcal G$ is spanned by feasible partial assignments.

\item \textit{Initialization.}
Prepare the root state $\ket{\psi(0)}=\ket{u}^{\otimes n}$.

\item \textit{Quantum evolution.}
Evolve under the time-dependent Hamiltonian $H(t)=U(t)H_{\rm P}U^\dagger(t)$ for a total time $T$. Choose the externally controlled variation of $U(t)$ to suppress transitions across the constraint gap $\Delta$.

\item \textit{Measurement.}
Apply $U^\dagger(T)$ and measure the qutrits in the computational basis to obtain a feasible partial assignment $\mu=(\mu_1,\ldots,\mu_n)$, where $\mu_i\in\{u,0,1\}$.

\item \textit{Classical postprocessing.}
If $D(\mu)=n$, return $x=\mu$ as an exact solution. Otherwise, retain every assigned value and independently assign each remaining variable uniformly from $\{0,1\}$ to obtain a complete Boolean assignment $x$.
\end{enumerate}

Adiabaticity with respect to the constraint gap $\Delta$ suppresses transitions between the instantaneous feasible manifold $\mathcal G(t)$ and the constraint-violating sector. The evolution in step 3 is therefore confined, to leading order, to $\mathcal G(t)$ and governed by the non-Abelian gauge matrix $A(t)$. If the secondary Hamiltonian $H_{\rm sec}(t)=-A(t)$ has a gap $\Delta_A$, one may additionally require $U(t)$ to vary slowly relative to $\Delta_A$, so that the state follows an instantaneous eigenstate of $-A(t)$. In typical protocols, $\Delta_A$ depends on $n$ and is smaller than $\Delta$, making this second condition more restrictive. Crucially, however, QPF does not require eigenstate following within the feasible manifold: nonadiabatic gauge dynamics and coherent interference are also admissible algorithmic resources.

The initialization step can be modified to incorporate a classically obtained feasible partial assignment, providing a natural warm start. For problems such as maximum-independent-set (MIS) in which the complete-solution manifold $\mathcal G_n$ is connected, one can instead initialize the system in a trivial solution(empty set) and restrict the local rotation to
\begin{equation}
U_i(t)
=
\begin{pmatrix}
1 & 0\\
0 & V_i(t)
\end{pmatrix},
\qquad V_i(t)\in SU(2).
\end{equation}
QPF then reduces to the non-Abelian mixing algorithms of Refs.~\cite{WuYuWilczek2020,YuWilczekWu2021}.

\subsection{Control protocols}
The performance of QPF depends on the time dependence of $U(t)$ or, equivalently, $A(t)$. For the gauge matrix derived in Sec.~\ref{sec:nagmx}, control is exercised through the three parameters $\theta(t)$, $\phi(t)$, and $\eta(t)$.

The connected feasible graph in Fig.~\ref{fig:feasible_layers}(c) can be viewed as an irregular lattice whose on-site energies are the diagonal elements $-A_{\mu\mu}$. Equation~\eqref{eq:gauge-diagonal} shows that the gauge potential introduces a depth-dependent energy bias. Following the maximum-independent-set construction~\cite{WuYuWilczek2020,YuWilczekWu2021}, our protocols vary $\theta$ slowly from $0$ to $\pi$ so that $A_{\mu\mu}$ initially favors the root configuration $\ket{u}^{\otimes n}$ and finally favors the maximal-depth sector. At each fixed $\theta$, the potential is monotonic in the assignment depth $D(\mu)$.

The diagonal matrix elements at the two endpoints are
\begin{subequations}
\label{eq:endpoint-diagonal}
\begin{align}
-A_{\mu\mu}(0)
&=
n\dot{\phi}(0)
+
D(\mu)
\left[
\frac{\dot{\eta}(0)}{2}
-
\dot{\phi}(0)
\right],
\label{eq:endpoint-diagonal-initial}
\\
-A_{\mu\mu}(T)
&=
\frac{D(\mu)}{2}
\left[
\dot{\phi}(T)+\dot{\eta}(T)
\right].
\label{eq:endpoint-diagonal-final}
\end{align}
\end{subequations}
The parameters should therefore be chosen to satisfy
\begin{equation}
\label{equ:gsreq}
\dot{\eta}(0)>2\dot{\phi}(0),
\qquad
\dot{\eta}(T)<-\dot{\phi}(T).
\end{equation}

The off-diagonal matrix elements govern coherent hopping on the feasible graph. The transitions $\ket{u}\leftrightarrow\ket{0},\ket{1}$ correspond to the single-site assignment flips introduced above and connect neighboring depth sectors, as illustrated in Fig.~\ref{fig:feasible_layers}(c). In addition, the qutrit gauge matrix generates the same-depth transition $\ket{0}\leftrightarrow\ket{1}$, which couples feasible configurations within the same depth sector. 

We introduce a same-depth control parameter
\begin{equation}
g(t)=\dot{\eta}(t)-\sin^2\frac{\theta(t)}{2}\dot{\phi}(t).
\label{eq:g-definition}
\end{equation}
A simple choice is $g=0$, for which the $\ket{0}\leftrightarrow\ket{1}$ matrix element vanishes and only single-site assignment flips remain. This gives a direct qutrit generalization of the MIS dynamics. For the schedules considered below, $\dot{\phi}<0$ ensures that this choice also satisfies the endpoint requirements. The diagonal matrix element reduces to the MIS form under the replacement $n_\mu\rightarrow D(\mu)$, where $n_\mu$ denotes the number of 1s in the MIS configuration. The off-diagonal hopping amplitudes differ from those in the MIS construction by an overall factor of $1/\sqrt{2}$. The nonzero elements of the restricted gauge matrix $A^{(g=0)}(t)$ are
\begin{align}
A_{\mu\mu}^{(g=0)}
&=
-\left[
(n-D(\mu))\cos^2\frac{\theta}{2}
+
D(\mu)\sin^2\frac{\theta}{2}
\right]\dot{\phi},
\\
A_{\mu\nu}^{(g=0)}
&=
\frac{1}{\sqrt{2}}
\left(
\frac{1}{2}\sin\theta\,\dot{\phi}
-
\frac{i}{2}\dot{\theta}
\right),
\quad
\mu_i=u,\;\nu_i\in\{0,1\},
\end{align}
with $A_{\nu\mu}^{(g=0)}=\left(A_{\mu\nu}^{(g=0)}\right)^*$ by Hermiticity.

In the remainder of this work, unless otherwise stated, we take $g(t)=0$ and use $A^{(g=0)}(t)$ as the gauge matrix governing the evolution. Specifically, following the two-rate protocol of Ref.~\cite{YuWilczekWu2021}, we adopt
\begin{equation}
\begin{aligned}
\theta(t)&=\omega_\theta t,\\
\dot{\phi}(t)&=-\omega_\phi,\\
\dot{\eta}(t)&=-\omega_\phi\sin^2\frac{\theta(t)}{2}\,,
\end{aligned}
\label{eq:algorithm-control-path}
\end{equation}
with $\omega_\theta\ll \omega_\phi$. 
The resulting non-Abelian transport through the feasible manifold is illustrated schematically in Fig.~\ref{fig:feasible_layers}(d).

\subsection{Spin-1 Schr\"odinger-picture Hamiltonian}
\label{sec:spin1-hamiltonian}

The non-Abelian gauge dynamics admits an equivalent fixed-basis Schr\"odinger-picture formulation. We introduce the rotating-frame state $\ket{\tilde\psi(t)}=U^\dagger(t)\ket{\psi(t)}$, under which the evolution generated by $H(t)=U(t)H_{\rm P}U^\dagger(t)$ is described by
\begin{equation}
H_{\rm S}(t)=H_{\rm P}+H_1(t),
\qquad
H_1(t)=-iU^\dagger(t)\dot U(t).
\label{eq:schrodinger-picture-H}
\end{equation}
In this representation, the constraint Hamiltonian $H_{\rm P}$ remains time independent, while $H_1(t)$ provides the local driving whose projection generates the non-Abelian dynamics within the feasible manifold.

Since $U(t)=\bigotimes_i U_i(t)$, the driving Hamiltonian is a sum of single-qutrit generators,
\begin{equation}
H_1(t)=-i\sum_i U_i^\dagger(t)\dot U_i(t).
\end{equation}
For the $g=0$ protocol, substituting Eq.~\eqref{eq:local-qutrit-unitary} gives, up to an irrelevant scalar term,
\begin{align}
H_{\rm S}(t)
&=H_{\rm P}
-\dot\phi(t)\cos\theta(t)\sum_i P_i
\nonumber\\
&\quad-\frac{\sin\theta(t)\dot\phi(t)}{2}\sum_i S_i^x
+\frac{\dot\theta(t)}{2}\sum_i\{S_i^y,S_i^z\},
\label{eq:schrodinger-H-general}
\end{align}
where $P_i=(S_i^z)^2$ and $\{A,B\}=AB+BA$ denotes the anticommutator. We denote the projector onto the local assigned subspace by $P_i$ to distinguish it from the search-space size $N=2^n$ introduced below.

The local driving terms have distinct physical roles. Since $\sum_iP_i\ket{\mu}=D(\mu)\ket{\mu}$, the first term introduces a depth-dependent on-site potential controlled by $-\dot\phi\cos\theta$. The remaining terms couple the unassigned state $\ket{u}$ to the assigned states $\ket{0}$ and $\ket{1}$, generating coherent hopping between neighboring assignment-depth sectors. In particular, the spin-$1$ quadrupolar operator $\{S_i^y,S_i^z\}$ provides a local control component absent in the qubit MIS construction. The constraint Hamiltonian $H_{\rm P}$ separates the feasible manifold $\mathcal G$ from the constraint-violating sector by an energy gap of at least $\Delta$. A conservative sufficient condition for maintaining spectral separation under local driving is
\begin{equation}
\Delta\gg\omega_{\rm drive}
=n\max_t\sqrt{\dot\phi(t)^2+\dot\theta(t)^2},
\label{eq:constraint-protection}
\end{equation}
where $\omega_{\rm drive}$ denotes the maximum instantaneous spectral width of $H_1(t)$, and the explicit expression applies to the homogeneous $g=0$ protocol.

In the strong-constraint regime, degenerate perturbation theory~\cite{BravyiDiVincenzoLoss2011} gives the leading-order effective Hamiltonian within $\mathcal G$,
\begin{equation}
H_{\rm sec}(t)=P_{\mathcal G}H_1(t)P_{\mathcal G}=-A(t),
\end{equation}
with $H_1(t)$ defined in Eq.~\eqref{eq:schrodinger-picture-H}. Finite constraint strength introduces corrections whose accumulated effects depend on the evolution time and control schedule. Constraint protection is therefore distinct from the adiabatic requirement associated with the secondary spectral gap $\Delta_A$.

\subsection{Superconducting-qutrit implementation}
\label{sec:experimental-realization}

Equation~\eqref{eq:schrodinger-H-general} expresses the dynamics in terms of static constraints and local driving. To illustrate a possible superconducting implementation, we consider the homogeneous $2$-XORSAT constraint $x_i\oplus x_j=0$, for which
\begin{equation}
H_{\rm P}=\frac{\Delta}{2}\sum_{\langle ij\rangle}
\left(P_iP_j-S_i^zS_j^z\right).
\label{eq:2xorsat-constraint-H}
\end{equation}
We encode $\ket{u}=\ket{g}$, $\ket{0}=\ket{e}$, and $\ket{1}=\ket{f}$ in the three lowest transmon levels~\cite{Koch2007Transmon,Morvan2021QutritRB}. Writing $n_{ir}=\ket{r}_i\bra{r}$ gives $P_i=n_{ie}+n_{if}$ and $S_i^z=n_{ie}-n_{if}$.

In a rotating frame, the required controls can be expressed as
\begin{align}
H_{\rm circ}(t)
&=\sum_i\sum_{r=e,f}\delta_{ir}(t)n_{ir}
\nonumber\\
&\quad+\frac{1}{2}\sum_i\sum_{r=e,f}
\left[\Omega_{ir}(t)\ket{g}_i\bra{r}+\mathrm{H.c.}\right]
\nonumber\\
&\quad+\sum_{\langle ij\rangle}\sum_{r,s=e,f}
\chi_{rs}^{ij}n_{ir}n_{js},
\label{eq:circuit-hamiltonian}
\end{align}
where $\delta_{ir}$ are effective detunings including local energy shifts, $\Omega_{ir}$ are complex coupling amplitudes, and $\chi_{rs}^{ij}$ are conditional energy shifts. The mapping to Eq.~\eqref{eq:schrodinger-H-general} is
\begin{align}
\delta_{ie}(t)=\delta_{if}(t)
&=-\dot\phi\cos\theta,
\nonumber\\
\Omega_{ie}(t)=\Omega_{if}(t)
&=\frac{-\dot\phi\sin\theta+i\dot\theta}{\sqrt{2}},
\nonumber\\
\chi_{ef}^{ij}=\chi_{fe}^{ij}&=\Delta,
\qquad
\chi_{ee}^{ij}=\chi_{ff}^{ij}=0.
\label{eq:circuit-parameter-mapping}
\end{align}
The interaction therefore penalizes $\ket{e_i f_j}$ and $\ket{f_i e_j}$, while leaving configurations containing an unassigned state $\ket{g}$ unpenalized.

Microwave control of transmons provides single-photon $\ket{g}\leftrightarrow\ket{e}$ and effective two-photon $\ket{g}\leftrightarrow\ket{f}$ couplings with tunable amplitudes and phases; dynamic phase corrections compensate the associated ac-Stark shifts~\cite{Vepsalainen2019Superadiabatic}. Conditional Stark shifts controlled by microwave amplitude, phase, and frequency, as well as flux-tunable couplers, provide experimentally demonstrated mechanisms for tuning qutrit cross-Kerr interactions~\cite{Goss2022QutritCrossKerr,Luo2023TwoQutritGate}. These results motivate a possible realization of the mapped Hamiltonian, with the specific conditional-shift pattern in Eq.~\eqref{eq:circuit-parameter-mapping} remaining a requirement for interaction engineering.

\section{Applications}
\label{sec:applications}

We now illustrate the framework through three representative problems. We first examine unstructured search, in which the oracle is a single constraint supported on all $n$ variables, and test whether the qutrit construction reproduces Grover scaling~\cite{Grover1997Search,FarhiGutmann1998Analog,ShenviKempeWhaley2003}. We then use $k$-XORSAT to study how local constraints shape the spectrum and dynamics of the feasible manifold. Finally, we formulate $k$-SAT within the same framework and evaluate the resulting complete assignments by the fraction of satisfied clauses.

\subsection{Unstructured search}
\label{sec:unstructured-search}

\begin{figure}[!t]
\centering
\includegraphics[width=\linewidth]{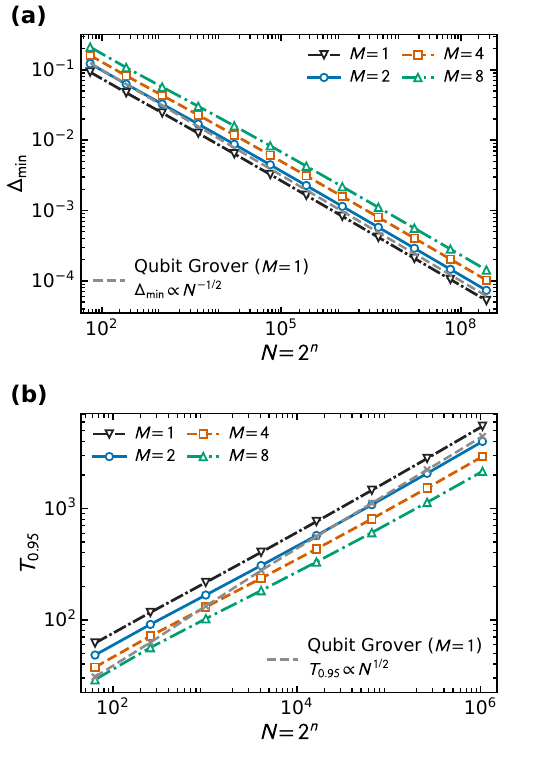}
\caption{
Grover search with ideal projected qutrit dynamics under a local adiabatic schedule for $N=2^n$ and $M=1,2,4,8$ marked states. (a) Inverse minimum effective gap in the accessible symmetry sector. (b) Runtime $T_{0.95}$ at the first numerically resolved crossing of $P_{\rm succ}=0.95$. Gray dashed curves give the single-target qubit Grover gap in (a) and the local-adiabatic runtime in (b), following Ref.~\cite{RolandCerf2002}.
}
\label{fig:grover}
\end{figure}

The original search problem is defined over $N=2^n$ binary strings, with a marked set $\mathcal{M}\subseteq\{0,1\}^n$ of size $|\mathcal{M}|=M$. The goal is to find one marked string. In the qutrit formulation introduced above, the original Boolean search space is the full-assignment subspace of the qutrit Hilbert space. The projector onto this sector and the projector onto the marked subspace are
\begin{align}
P_{\rm full}
&=
\prod_{i=1}^{n}
\left(
\ket{0}_i\bra{0}
+
\ket{1}_i\bra{1}
\right),
\\
P_{\mathcal M}
&=
\sum_{x\in\mathcal M}
\ket{x}\bra{x}.
\end{align}

We choose the search Hamiltonian as
\begin{equation}
H_{\rm search}
=
\Delta
\left(
P_{\rm full}
-
P_{\mathcal M}
\right),
\label{eq:search-hamiltonian}
\end{equation}
which is positive semidefinite since $P_{\mathcal M}\leq P_{\rm full}$. For any partial configuration with $D(\mu)<n$, both projectors vanish. For a complete configuration $\ket{x}$, $H_{\rm search}\ket{x}=0$ if $x\in\mathcal M$ and $H_{\rm search}\ket{x}=\Delta\ket{x}$ otherwise. Thus, all partial configurations with $D<n$ belong to the feasible manifold, whereas the complete-assignment sector contains only the marked states, giving
\begin{equation}
    \dim\mathcal G_D
    =
    \begin{cases}
        \binom{n}{D}2^D, & D<n,\\[4pt]
        M, & D=n.
    \end{cases}
    \label{eq:search-layer-size}
\end{equation}
Hence the oracle leaves every intermediate layer unrestricted and acts only at the complete-assignment boundary. We take the $g=0$ protocol with $\dot\phi=-\omega$ ($\omega\equiv\omega_\phi>0$) and sweep $\theta(t)$ from $0$ to $\pi$. Freezing $\theta$ and omitting the term proportional to $\dot\theta$, Eq.~\eqref{eq:projected-effective-H} gives, up to a scalar shift,
\begin{equation}
    H_{\rm eff}^{\rm Grover}(\theta)
    =
    \omega P_{\mathcal G}
    \left[
        \cos\theta\sum_i P_i
        +
        \frac{\sin\theta}{2}\sum_i S_i^x
    \right]
    P_{\mathcal G}.
    \label{eq:search-gauge-H}
\end{equation}

The search bottleneck is most transparent in the layer representation. Within the incomplete sector, the uniform layer states
\begin{equation}
    \ket D =
    \frac{1}{\sqrt{2^D\binom{n}{D}}}
    \sum_{\mu\in\mathcal G_D}\ket{\mu},
    \qquad D=0,\ldots,n-1,
\end{equation}
span a symmetric chain with on-site energies $\omega D\cos\theta$ and nearest-neighbor couplings $\frac{\omega\sin\theta}{2}\sqrt{(n-D)(D+1)}$. The search constraint enters at the final boundary, where the unrestricted complete-assignment layer is replaced by $M$ marked states. Since each marked configuration has $n$ predecessors, the coupling to the uniform marked state $\ket{\psi_{\rm M}}=M^{-1/2}\sum_{x\in\mathcal M}\ket x$ is
\begin{equation}
\begin{aligned}
    \bra{\psi_{\rm M}} H_{\rm eff}^{\rm Grover}\ket{n-1}
    &=
    \frac{\omega n\sin\theta}{2\sqrt{2}}
    \sqrt{\frac{\dim\mathcal G_n}{\dim\mathcal G_{n-1}}}
    \\
    &=
    \frac{\omega\sin\theta}{2}\sqrt{\frac{nM}{N}}.
\end{aligned}
\end{equation}
This suggests an avoided crossing between the incomplete-sector branch and a dressed marked branch, analogous to the final-link bottleneck in quantum optimization on flat energy landscapes~\cite{Cain2023FlatLandscapes}. Under the isolated-crossing and finite-overlap assumptions specified in Appendix~\ref{app:grover-gap}, with $n\cos^2(\theta_c/2)=\Theta(1)$ at resonance, the gap scales as
\begin{equation}
    \Delta_{\rm cross}
    =
    \Theta\!\left(\omega\sqrt{\frac{M}{N}}\right).
    \label{eq:grover-gap-scaling}
\end{equation}

For fixed $M$, this scaling is consistent with Fig.~\ref{fig:grover}(a), where the inverse minimum gap follows the expected $\sqrt N$ behavior over the system sizes studied. Within this effective two-level description, a local adiabatic schedule~\cite{RolandCerf2002} yields
\begin{equation}
    T_{\rm qutrit}
    =
    O\!\left[
        \frac{\operatorname{poly}(n)}{\omega}
        \sqrt{\frac{N}{M}}
    \right],
    \label{eq:qutrit-grover-runtime}
\end{equation}
recovering the quadratic Grover scaling up to polynomial factors in $n=\log_2N$. Since the gap alone does not determine the finite-time success probability, we also solve the full time-dependent Schr\"odinger equation. Figure~\ref{fig:grover}(b) shows that the runtime required to directly measure a marked state with probability $P_{\rm succ}=0.95$ exhibits the expected Grover-type scaling over the accessible system sizes. 

\subsection{\texorpdfstring{$k$-XORSAT}{k-XORSAT}}

We next consider $k$-XORSAT, in which each constraint fixes the parity of $k$ Boolean variables,
\begin{equation}
    \bigoplus_{j\in\partial\alpha}x_j=b_\alpha,
    \qquad
    b_\alpha\in\{0,1\},
    \label{eq:xorsat-constraint}
\end{equation}
where $\partial\alpha$ is the set of variables involved in constraint $C_\alpha$, with $|\partial\alpha|=k$. For $k<n$, these constraints can exclude configurations at intermediate assignment depths, as soon as all variables in a constraint have been assigned. Although $k$-XORSAT can be solved in polynomial time by Gaussian elimination over $\mathrm{GF}(2)$~\cite{AchlioptasMolloy2015}, it provides a simple setting for studying how local constraints shape the feasible manifold and the associated gauge dynamics.

Using the qutrit encoding in Eq.~\eqref{eq:qutrit-spin1-encoding}, the problem Hamiltonian is
\begin{equation}
    H_{k\text{-XORSAT}}
    =
    \frac{\Delta}{2}
    \sum_\alpha
    \left[
        \prod_{j\in\partial\alpha}(S_j^z)^2
        -
        (-1)^{b_\alpha}
        \prod_{j\in\partial\alpha}S_j^z
    \right].
    \label{eq:xorsat-problem-hamiltonian}
\end{equation}
Each constraint contributes an energy penalty $\Delta$ only when all variables in its support are assigned and their parity disagrees with $b_\alpha$. The zero-energy manifold of $H_{k\text{-XORSAT}}$ is therefore spanned by all partial assignments that violate no constraint.

We now specialize to satisfiable complete $3$-XORSAT instances with $n\geq4$, defined by a parity constraint on every triple,
\begin{equation}
    x_i\oplus x_j\oplus x_k=b_{ijk},
    \qquad
    1\leq i<j<k\leq n.
\end{equation}
For any satisfiable instance, the relabeling $x_i\mapsto x_i\oplus z_i$, with $z$ a satisfying assignment, maps all constraints to $x_i\oplus x_j\oplus x_k=0$ without changing the $g=0$ gauge spectrum. We therefore consider the homogeneous instance below. In this representation, all assigned variables must be zero for $D\geq4$. Let $|n_0,n_1\rangle$ denote the normalized uniform superposition of configurations with $n_0$ assigned zeros and $n_1$ assigned ones. The permutation-invariant feasible subspace is spanned by the assignment-depth chain
\begin{equation}
    |D\rangle\equiv|D,0\rangle,
    \qquad D=0,\ldots,n,
\end{equation}
and four additional states, $|0,1\rangle$, $|1,1\rangle$, $|0,2\rangle$, and $|1,2\rangle$, confined to $D\leq3$. The homogeneous reduction and orbit classification are derived in Appendix~\ref{app:3xor-gap}.

\begin{figure}[!t]
\centering
\includegraphics[width=\linewidth]{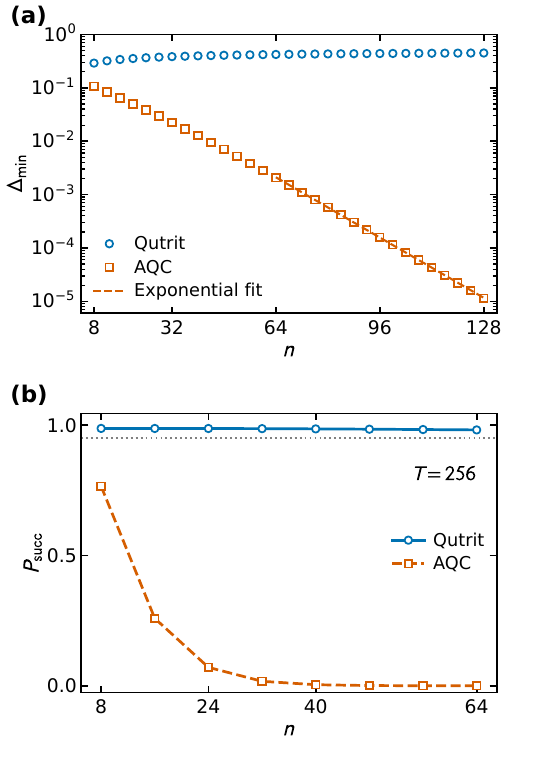}
\caption{
Spectral and dynamical comparison for complete $3$-XORSAT. (a) Minimum gaps in the permutation-invariant sectors of the frozen $g=0$ qutrit gauge Hamiltonian and the conventional transverse-field AQC Hamiltonian. The dashed line shows the fit $\Delta_{\min}^{\mathrm{AQC}}/\omega
\simeq0.382e^{-0.0812n}$ over $64\leq n\leq128$.
(b) Final exact-solution probability at $\omega T=256$, using the linear schedules $\theta(t)=\pi t/T$ and $s(t)=t/T$. The horizontal dotted line marks $P_{\mathrm{succ}}=0.95$. The qutrit dynamics retain the term proportional to $\dot\theta$ and assume ideal projection onto the feasible subspace. The qutrit gap in (a) is the projected gauge gap, distinct from the physical constraint gap.
}
\label{fig:3xorsat}
\end{figure}

For the frozen $g=0$ gauge Hamiltonian, with $\dot\phi=-\omega$ and $\dot\theta=0$, the chain Hamiltonian is equivalent to a collective spin $J=n/2$,
\begin{equation}
    \frac{H_{\rm ch}}{\omega}
    =
    \frac n2\cos\theta
    +
    \cos\theta\,J_z
    -
    \frac{\sin\theta}{\sqrt2}J_x.
\end{equation}
It describes a single large spin in a nonvanishing effective field and therefore has the size-independent level spacing
\begin{equation}
    \Delta_{\rm ch}(\theta)
    =
    \omega\sqrt{\cos^2\theta+\frac12\sin^2\theta}
    \geq\frac{\omega}{\sqrt2}.
\end{equation}
The four shallow orbit states couple to the chain but do not close the gap. Appendix~\ref{app:3xor-gap} establishes a uniform, size-independent lower bound for the projected gauge gap within the permutation-invariant sector. This constant-gap behavior is also observed numerically in Fig.~\ref{fig:3xorsat}(a).

We compare the qutrit protocol with the conventional transverse-field adiabatic Hamiltonian,
\begin{equation}
    H_{\rm AQC}(s)=(1-s)H_X+sH_Z,
    \qquad 0\leq s\leq1,
    \label{eq:xorsat-aqc-H}
\end{equation}
where $H_X$ is a uniform transverse-field driver and $H_Z$ penalizes violated parity constraints of the homogeneous complete $3$-XORSAT instance. Both terms are normalized to an extensive energy scale $O(n\omega)$. This path undergoes a first-order quantum phase transition with an exponentially small minimum gap~\cite{Jorg2010Gap,OhkuwaNishimori2017}. Figure~\ref{fig:3xorsat}(a) compares this exponential gap closing with the size-independent qutrit gauge gap. At fixed runtime $\omega T=256$, the qutrit protocol maintains an exact-solution probability close to unity over the system sizes studied, while the AQC success probability decreases rapidly with $n$ [Fig.~\ref{fig:3xorsat}(b)].

\subsection{\texorpdfstring{Approximation for $k$-SAT}{Approximation for k-SAT}}
\label{sec:approximation-ksat}

We demonstrate that non-Abelian mixing can also serve as an approximation algorithm for hard constraint-satisfaction problems, taking $k$-SAT as an example. A $k$-SAT formula consists of $m$ clauses,
\begin{equation}
    F=\bigwedge_{\alpha=1}^{m}C_\alpha,
    \qquad
    C_\alpha=\bigvee_{r=1}^{k}\ell_{\alpha r},
    \label{eq:ksat-clause}
\end{equation}
where each literal $\ell_{\alpha r}$ is either a Boolean variable $x_{i_{\alpha r}}$ or its negation $\neg x_{i_{\alpha r}}$. A complete assignment satisfies the formula if every clause contains at least one true literal.

To construct the qutrit Hamiltonian, we associate a sign $s_{\alpha r}=+1$ with a positive literal and $s_{\alpha r}=-1$ with a negated literal. Under the encoding $|0\rangle_i\equiv|S_i^z=+1\rangle$ and $|1\rangle_i\equiv|S_i^z=-1\rangle$, the projector onto the Boolean value that makes the literal false is $[(S_{i_{\alpha r}}^z)^2+s_{\alpha r}S_{i_{\alpha r}}^z]/2$. Since a clause is violated only when all of its literals are false, the problem Hamiltonian is
\begin{equation}
    H_{k\text{-SAT}}
    =
    \frac{\Delta}{2^k}
    \sum_{\alpha=1}^{m}
    \prod_{r=1}^{k}
    \left[
        (S_{i_{\alpha r}}^z)^2
        +
        s_{\alpha r}S_{i_{\alpha r}}^z
    \right].
    \label{eq:ksat-problem-hamiltonian}
\end{equation}
Each clause term is $\Delta$ times the projector onto the local Boolean assignment that makes all literals in that clause false. Since $S_i^z|u\rangle_i=0$, a clause contributes an energy penalty only when all of its variables are assigned and all its literals are false. Hence the zero-energy manifold consists precisely of the partial assignments that have not yet violated any clause.

The approximation quality is not determined by the assignment depth alone. We denote by $N_{\rm sat}(\,\cdot\,)$ the number of clauses already satisfied by a complete or partial assignment and define
\begin{equation}
    Q(\,\cdot\,)=\frac{N_{\rm sat}(\,\cdot\,)}{m},
    \qquad
    Q^\star=\max_{x\in\{0,1\}^n}Q(x).
\end{equation}
Thus $Q(x)$ gives the satisfied-clause fraction for a complete assignment, whereas $Q(\mu)$ gives the fraction of clauses already satisfied by a partial assignment. For the satisfiable instances considered below, $Q^\star=1$.

\begin{figure}[!t]
\centering
\includegraphics[width=\columnwidth]{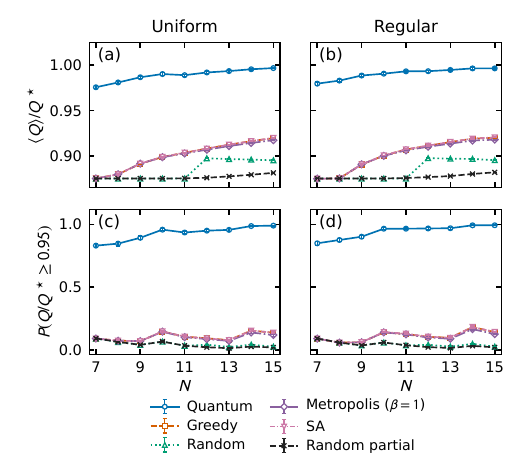}
\caption{
Finite-size performance for satisfiable $3$-SAT at $m/n=6$. (a),(b) Mean normalized objective and (c),(d) probability of reaching $95\%$ of the optimum for uniform and degree-$18$ regular ensembles, respectively. Quantum evolution uses $T=n^2$; all classical baselines use $B=n^2$ clause inspections. SA uses a schedule selected on separate training instances. Random partial assignment is budget-truncated, followed by uniform completion, with final scoring included in the budget. Points average $30$ shared instances; error bars denote one standard error. Stochastic classical estimates use $4\,096$ runs per instance; the uniform-random baseline and optimum are evaluated exactly. Quantum time and classical inspections are distinct resources; budget dependence is reported in Appendix~\ref{app:classical-budget}.
}
\label{fig:max3sat}
\end{figure}

To assess the approximation performance of the non-Abelian dynamics, we benchmark the qutrit protocol on satisfiable $3$-SAT instances drawn from random signed-clause and degree-$18$ regular ensembles at clause density $m/n=6$. The quantum results are obtained from ideal projected evolution with runtime $T=n^2$, followed by computational-basis measurement and uniform completion of any remaining unassigned variables. Performance is evaluated on the resulting complete assignments. The classical baselines include greedy search, uniform random sampling, Metropolis sampling with $\beta=1$~\cite{Metropolis1953}, simulated annealing~\cite{Kirkpatrick1983}, and random partial assignment, each subject to a budget of $n^2$ clause inspections. The budget accounting and dependence of the results on the classical budget are discussed in Appendix~\ref{app:classical-budget}.

Across both ensembles, the qutrit protocol produces normalized objective values close to unity, with $\langle Q\rangle/Q^\star\gtrsim0.97$ over the system sizes studied [Figs.~\ref{fig:max3sat}(a) and (b)]. The classical baselines remain systematically lower over the same range. The separation is more pronounced for the probability of obtaining a near-optimal assignment: $P(Q/Q^\star\geq0.95)$ remains high for the qutrit protocol but is substantially smaller for the classical methods [Figs.~\ref{fig:max3sat}(c) and (d)].

These finite-size results support non-Abelian feasible-manifold transport as an approximation strategy for $3$-SAT. Because quantum evolution time and classical clause-inspection counts measure different resources, however, this comparison is a finite-size performance benchmark rather than evidence of an equal-cost or asymptotic quantum advantage.

\section{Discussion}
\label{sec:discussion}
QPF occupies a middle ground between several familiar pictures of quantum dynamics while retaining a distinct algorithmic logic. Like Hamiltonian computation, it is formulated through continuous-time evolution; like a quantum walk, it generates transport on a graph of computational configurations; and like constrained-mixer methods, it uses the structure of the feasible subspace~\cite{FarhiGutmann1998DecisionTree,Aharonov2001QuantumWalks,Kempe2003QuantumWalks,Childs2003GluedTrees,ChildsGoldstone2004,HenSpedalieri2016,HenSarandy2016,Hadfield2019,Fuchs2022Mixers}. Its distinguishing feature is that this graph includes feasible partial assignments and arises as the degenerate manifold of an isospectrally rotated problem Hamiltonian. The non-Abelian gauge potential supplies a systematic connection between local physical controls and coherent motion on that graph. Describing QPF as a separate paradigm is therefore a statement about algorithm design, not a claim that it lies outside the circuit model: any finite-time QPF evolution can, in principle, be approximated by a quantum circuit. Conversely, the availability of a circuit compilation does not eliminate the conceptual and practical value of designing the dynamics at the level of paths, constraints, and gauge fields.

\subsection{Analogy with nondeterministic Turing machines}

The path-finding picture invites comparison with a nondeterministic Turing machine (NTM), but the two must be distinguished carefully. An NTM may have several allowed transitions from a configuration. Its computation is represented by a tree of possible branches, and a decision problem is accepted when at least one branch accepts~\cite{Sipser2013,AroraBarak2009}. In QPF, assignment moves likewise generate many possible paths from the root $\ket u^{\otimes n}$ through partial configurations toward complete assignments. The feasible graph provides a geometric representation of these possibilities, and a satisfying assignment is reached through one or more connected paths.

The quantum state, however, is not a collection of independently executed classical branches. It is a single coherent superposition over feasible configurations, with a complex amplitude for each configuration. Different paths can interfere constructively or destructively, and the controls modify all amplitudes through one unitary evolution. Measurement then samples an outcome; it does not reveal whether an accepting branch exists somewhere in a computation tree. In contrast to the existential acceptance rule of an NTM, QPF requires the controlled dynamics to place sufficient probability on useful outcomes. The analogy is therefore structural rather than a complexity-theoretic equivalence: nondeterminism supplies a useful vocabulary for branching computation, whereas QPF supplies coherent, phase-sensitive navigation of a related configuration space. In particular, the construction does not imply that NP problems can generally be solved efficiently by quantum evolution, because the feasible graph may be exponentially large and the probability of reaching a solution may remain exponentially small.

This comparison also clarifies the role of algorithm design. For an NTM, the transition relation specifies which branches are allowed; for QPF, one must additionally design amplitudes and phases so that paths leading to deep or complete assignments are favored. The depth-dependent diagonal gauge potential biases transport across layers, while off-diagonal terms determine connectivity and interference. Same-depth mixing, controlled by $g(t)$, can reorganize amplitudes within a layer, and warm starts can restrict the initial region of the graph. Future work can therefore draw on optimal control, shortcut-to-adiabaticity methods, and problem-specific graph structure without assuming adiabatic eigenstate following within $\mathcal G$.

\subsection{Opportunities, limitations, and open questions}

The separation of the constraint gap $\Delta$ from the secondary gap $\Delta_A$ creates two levels of control. Constraint protection maintains the computational manifold, whereas the secondary dynamics may be adiabatic, deliberately nonadiabatic, or optimized for a finite-time target. This flexibility is a principal advantage over a conventional ground-state interpolation, but it also shifts responsibility to schedule design. General performance bounds will require assumptions about the feasible graph, spectral structure, and control path; a large constraint gap alone says nothing about the probability of reaching a solution.

Several limitations define the near-term research agenda. First, the numerical application studies use ideal projected evolution, so leakage, control errors, decoherence, and measurement errors remain to be quantified. Second, the sufficient constraint-protection condition in Eq.~\eqref{eq:constraint-protection} contains an extensive drive scale for homogeneous local control, and finite evolution times introduce corrections beyond the leading projected Hamiltonian. Third, although the qutrit lifting preserves constraint support, arbitrary constraints may require high-body interactions or large operator expansions. The superconducting implementation outlined here is consequently a platform-specific route, not a universal hardware prescription. Fourth, the qutrit encoding and readout overheads must be included in resource estimates alongside evolution time.

The scope of the present evidence is correspondingly specific. Grover-type scaling is derived under explicit crossing and overlap assumptions and tested over accessible finite sizes. The constant-gap proof applies to the accessible permutation-invariant sector of complete $3$-XORSAT, not to general XORSAT or SAT instances. The $3$-SAT results are finite-size benchmarks under unequal quantum and classical resource measures, and sufficiently long classical runs can outperform the fixed quantum reference. Establishing broader advantage will require end-to-end resource accounting, comparisons with optimized classical algorithms, and scaling analyses beyond the symmetry-reduced regimes studied here. Open questions include which constraint graphs support scalable transport, how to choose nonadiabatic schedules systematically, when same-depth mixing improves success probabilities, and whether partial-assignment dynamics offers advantages for approximation or warm-started optimization beyond exact search.

\section{Conclusion}
\label{sec:conclusion}

We have introduced the Quantum Path Finder as a Hamiltonian algorithm-design paradigm based on coherent navigation of a degenerate manifold of feasible partial assignments. The qutrit encoding makes this manifold connected under local assignment moves, while local rotations generate non-Abelian gauge dynamics with tunable on-site potentials and hopping amplitudes. Constraint-protected evolution replaces the requirement of following a single adiabatic eigenstate path; adiabatic motion within the feasible manifold is an optional special case rather than the defining mechanism.

The framework recovers Grover-type search scaling under stated assumptions, yields a uniform gap for the accessible sector of complete $3$-XORSAT, and provides a finite-size approximation strategy for $3$-SAT. Together with the spin-$1$ formulation and proposed superconducting-qutrit mapping, these results establish a general route from Boolean constraints to controllable quantum paths. The central challenge now is to identify problem structures and control protocols for which this path-based paradigm delivers scalable, experimentally robust advantages under equal-resource comparisons.

\begin{acknowledgments}
 We thank Hongye Yu for helpful discussion. This work was supported by the National Natural Science Foundation of China (92365202, 12475011, and 11921005), the National Key R\&D Program of China (2024YFA1409002), and the Shanghai Municipal Science and Technology Project (25LZ2601100). The authors acknowledge ChatGPT and Kimi for assistance with numerical calculations, analytical derivations, and manuscript preparation. All AI-assisted results were independently verified, and the authors take full responsibility for this work.
\end{acknowledgments}
\appendix

\section{Derivation of the qutrit gauge matrix}
\label{app:gauge-derivation}

To derive the gauge matrix, we first introduce the symmetric and antisymmetric states
\begin{equation}
\ket{\pm}_i
=
\frac{\ket{0}_i\pm\ket{1}_i}{\sqrt{2}}.
\label{eq:app-pm-basis}
\end{equation}
In the basis $\{\ket{u}_i,\ket{+}_i,\ket{-}_i\}$, the local qutrit rotation takes the form
\begin{equation}
U_i(t)
=
\begin{pmatrix}
-e^{i\phi(t)}\cos\frac{\theta(t)}{2}
&
e^{i\phi(t)}\sin\frac{\theta(t)}{2}
&
0
\\[4pt]
\sin\frac{\theta(t)}{2}
&
\cos\frac{\theta(t)}{2}
&
0
\\[4pt]
0
&
0
&
e^{i\eta(t)}
\end{pmatrix}.
\label{eq:app-local-qutrit-unitary}
\end{equation}
The corresponding time-dependent local basis states are
\begin{subequations}
\label{eq:app-local-rotation}
\begin{align}
\ket{u(t)}_i
&=
-e^{i\phi(t)}
\cos\frac{\theta(t)}{2}\ket{u}_i
+
\sin\frac{\theta(t)}{2}\ket{+}_i,
\\
\ket{+(t)}_i
&=
e^{i\phi(t)}
\sin\frac{\theta(t)}{2}\ket{u}_i
+
\cos\frac{\theta(t)}{2}\ket{+}_i,
\\
\ket{-(t)}_i
&=
e^{i\eta(t)}\ket{-}_i.
\end{align}
\end{subequations}

Equivalently, transforming back to the computational basis $\{\ket{u}_i,\ket{0}_i,\ket{1}_i\}$ using
\begin{equation}
\ket{+}_i=\frac{\ket{0}_i+\ket{1}_i}{\sqrt{2}},
\qquad
\ket{-}_i=\frac{\ket{0}_i-\ket{1}_i}{\sqrt{2}},
\end{equation}
gives the local qutrit rotation in Eq.~\eqref{eq:local-qutrit-unitary} of the main text.

The projected non-Abelian gauge matrix is
\begin{equation}
A(t)
=
iP_{\mathcal G}U^\dagger(t)\dot U(t)P_{\mathcal G},
\label{eq:app-projected-gauge}
\end{equation}
where $P_{\mathcal G}$ projects onto the feasible manifold $\mathcal G$. Since $U(t)=\bigotimes_i U_i(t)$ is a product of local unitaries, its gauge generator decomposes as
\begin{equation}
iU^\dagger(t)\dot U(t)
=
\sum_i a_i(t),
\qquad
a_i(t)
=
iU_i^\dagger(t)\dot U_i(t).
\label{eq:app-local-gauge-generator}
\end{equation}

A direct calculation gives the local gauge generator in the basis $\{\ket{u}_i,\ket{+}_i,\ket{-}_i\}$ as
\begin{equation}
a_i(t)
=
\begin{pmatrix}
-\cos^2\frac{\theta}{2}\dot\phi &
\frac{1}{2}\sin\theta\,\dot\phi-\frac{i}{2}\dot\theta &
0
\\[4pt]
\frac{1}{2}\sin\theta\,\dot\phi+\frac{i}{2}\dot\theta &
-\sin^2\frac{\theta}{2}\dot\phi &
0
\\[4pt]
0 & 0 & -\dot\eta
\end{pmatrix},
\label{eq:app-local-gauge-plus-minus}
\end{equation}
where the explicit time dependence of $\theta(t)$, $\phi(t)$, and $\eta(t)$ has been suppressed for notational simplicity.

Transforming back to the computational basis $\{\ket{u}_i,\ket{0}_i,\ket{1}_i\}$ gives
\begin{align}
a_i(t)
&=
-\cos^2\frac{\theta}{2}\dot\phi\,
\ket{u}_i\bra{u}
\nonumber\\
&\quad
-
\frac{1}{2}
\left(
\sin^2\frac{\theta}{2}\dot\phi+\dot\eta
\right)
\left(
\ket{0}_i\bra{0}
+
\ket{1}_i\bra{1}
\right)
\nonumber\\
&\quad
+
\frac{1}{2}
\left(
\dot\eta-\sin^2\frac{\theta}{2}\dot\phi
\right)
\left(
\ket{0}_i\bra{1}
+
\ket{1}_i\bra{0}
\right)
\nonumber\\
&\quad
+
\frac{1}{\sqrt{2}}
\left(
\frac{1}{2}\sin\theta\,\dot\phi
-
\frac{i}{2}\dot\theta
\right)
\ket{u}_i
\left(
\bra{0}+\bra{1}
\right)_i
\nonumber\\
&\quad
+
\frac{1}{\sqrt{2}}
\left(
\frac{1}{2}\sin\theta\,\dot\phi
+
\frac{i}{2}\dot\theta
\right)
\left(
\ket{0}+\ket{1}
\right)_i
\bra{u}_i .
\label{eq:app-local-gauge-computational}
\end{align}

For a computational-basis state
\begin{equation}
\ket{\mu}
=
\ket{\mu_1}\otimes\cdots\otimes\ket{\mu_n},
\qquad
\mu_i\in\{u,0,1\},
\end{equation}
the projected gauge matrix is obtained by restricting $\sum_i a_i(t)$ to the feasible manifold $\mathcal G$.

For a feasible partial assignment $\mu$, each unassigned site contributes $-\cos^2(\theta/2)\dot\phi$, while each assigned site contributes $-[\sin^2(\theta/2)\dot\phi+\dot\eta]/2$. Since $\mu$ contains $n-D(\mu)$ unassigned sites and $D(\mu)$ assigned sites, the diagonal matrix element is
\begin{equation}
A_{\mu\mu}(t)
=
-(n-D(\mu))
\cos^2\frac{\theta}{2}\dot\phi
-
\frac{D(\mu)}{2}
\left(
\sin^2\frac{\theta}{2}\dot\phi
+
\dot\eta
\right).
\label{eq:app-gauge-diagonal}
\end{equation}

For two distinct feasible partial assignments $\mu$ and $\nu$, the off-diagonal matrix element can be nonzero only when they differ at a single site $i$, because the gauge generator is a sum of single-site operators. Reading the corresponding matrix elements from Eq.~\eqref{eq:app-local-gauge-computational}, we obtain
\begingroup\small
\begin{equation}
A_{\mu\nu}(t)
=
\setlength{\arraycolsep}{3pt}
\begin{cases}
\dfrac{1}{\sqrt{2}}
\left(
\dfrac{1}{2}\sin\theta\,\dot\phi
-
\dfrac{i}{2}\dot\theta
\right),
&
\mu_i=u,\quad \nu_i\in\{0,1\},
\\[10pt]
\dfrac{1}{\sqrt{2}}
\left(
\dfrac{1}{2}\sin\theta\,\dot\phi
+
\dfrac{i}{2}\dot\theta
\right),
&
\nu_i=u,\quad \mu_i\in\{0,1\},
\\[10pt]
\dfrac{1}{2}
\left(
\dot\eta
-
\sin^2\frac{\theta}{2}\dot\phi
\right),
&
\{\mu_i,\nu_i\}=\{0,1\},
\\[6pt]
0,
&
\text{otherwise}.
\end{cases}
\label{eq:app-gauge-offdiagonal}
\end{equation}
\endgroup

Equations~\eqref{eq:app-gauge-diagonal} and \eqref{eq:app-gauge-offdiagonal} reproduce the matrix elements quoted in the main text. The first two off-diagonal terms correspond to the depth-changing transitions $\ket{u}\leftrightarrow\ket{0},\ket{1}$, whereas the third corresponds to the same-depth transition $\ket{0}\leftrightarrow\ket{1}$. With the convention used here, the secondary Hamiltonian governing the evolution within the feasible manifold is $-A(t)$.

\section{More Hamiltonian instances}
\label{app:hamiltonian-instances}
\begin{table}[!htbp]
\caption{
Qutrit violation projectors.
}
\label{tab:boolean-qutrit-lifting}
\centering
\small
\setlength{\tabcolsep}{4pt}
\resizebox{\columnwidth}{!}{%
\begin{tabular}{@{}ll@{}}
\hline\hline
\textbf{Constraint}
&
\textbf{Qutrit}
\\
\hline

Unstructured search
&
$\displaystyle
\prod_{i=1}^{n}(S_i^z)^2
-\sum_{x\in\mathcal M}|x\rangle\langle x|
$
\\[12pt]

$k$-XORSAT
&
$\displaystyle
\frac{1}{2}
\left[
\prod_{i=1}^{k}(S_i^z)^2
-
(-1)^b\prod_{i=1}^{k}S_i^z
\right]
$
\\[12pt]

$k$-SAT
&
$\displaystyle
\prod_{i=1}^{k}
\frac{(S_i^z)^2+s_iS_i^z}{2}
$
\\[12pt]

Graph $2$-coloring
&
$\displaystyle
\frac{(S_i^z)^2(S_j^z)^2+S_i^zS_j^z}{2}
$
\\[12pt]

Equality, $x_i=x_j$
&
$\displaystyle
\frac{(S_i^z)^2(S_j^z)^2-S_i^zS_j^z}{2}
$
\\[12pt]

Implication, $x_i\Rightarrow x_j$
&
$\displaystyle
\frac{(S_i^z)^2-S_i^z}{2}
\frac{(S_j^z)^2+S_j^z}{2}
$
\\[12pt]

Independent set (edge)
&
$\displaystyle
\frac{(S_i^z)^2-S_i^z}{2}
\frac{(S_j^z)^2-S_j^z}{2}
$
\\[12pt]

NAE-$k$-SAT
&
$\displaystyle
\prod_{i=1}^{k}\frac{(S_i^z)^2+S_i^z}{2}
+
\prod_{i=1}^{k}\frac{(S_i^z)^2-S_i^z}{2}
$
\\[12pt]

$1$-in-$k$ SAT
&
$\displaystyle
\prod_{i=1}^{k}(S_i^z)^2
-
\sum_{j=1}^{k}
\frac{(S_j^z)^2-S_j^z}{2}
\prod_{\substack{i=1\\i\neq j}}^{k}
\frac{(S_i^z)^2+S_i^z}{2}
$
\\[12pt]

Cardinality, $|x|=r$
&
$\displaystyle
\sum_{\substack{x\in\{0,1\}^k\\|x|\neq r}}
\prod_{i=1}^{k}
\frac{(S_i^z)^2+(-1)^{x_i}S_i^z}{2}
$
\\
\hline\hline
\end{tabular}%
}
\end{table}

Table~\ref{tab:boolean-qutrit-lifting} summarizes the encodings used in the main text and several additional examples. Here $\mathcal M$ denotes the marked set; $b\in\{0,1\}$ is the required parity; and $s_i=+1$ ($-1$) denotes a positive (negated) literal. For independent set, $x_i=1$ denotes a selected vertex. The NAE-$k$-SAT and $1$-in-$k$ SAT entries assume positive literals; NAE means not-all-equal. For the cardinality constraint, $|x|=\sum_{i=1}^k x_i$ and $r\in\{0,\ldots,k\}$. Each lifted projector reproduces the Boolean violation penalty on complete assignments and vanishes whenever any participating variable is unassigned. A positive sum of these projectors therefore has all nonviolating partial assignments as zero-energy ground states. The lifting preserves the constraint support: a constraint on $k$ Boolean variables remains at most $k$-local on the qutrits, although its polynomial degree in $S_i^z$ can reach $2k$. General constraints may nevertheless require many-body interactions or large operator expansions, so an efficient physical implementation is not guaranteed.

\section{Grover scaling under the two-level approximation}
\label{app:grover-gap}

We consider the frozen Hamiltonian $H_{\rm eff}^{\rm Grover}(\theta)$ within the feasible sector, containing all incomplete configurations and a marked set $\mathcal M$ of size $M$, with $N=2^n$.

\subsection{Boundary coupling and avoided-crossing gap}

The uniform layer states $\ket D$ defined in the main text form a chain within the incomplete sector. Let $\ket{\psi_{\rm ch}}$ be its normalized lowest eigenstate, with energy $E_{\rm ch}$, and define the uniform marked state
\begin{equation}
    \ket{\psi_{\rm M}}
    =
    \frac{1}{\sqrt M}\sum_{x\in\mathcal M}\ket x.
\end{equation}
Only the final incomplete layer $\mathcal G_{n-1}$ couples directly to the marked sector. This layer contains $n2^{n-1}$ configurations, and each marked configuration has $n$ predecessors. The boundary coupling is therefore
\begin{equation}
\begin{aligned}
    \gamma
    &\equiv
    \left|
    \bra{\psi_{\rm M}}
    H_{\rm eff}^{\rm Grover}
    \ket{\psi_{\rm ch}}
    \right|\\
    &=
    \frac{\omega\sin\theta}{2}
    \sqrt{\frac{nM}{N}}\,
    \left|\braket{n-1|\psi_{\rm ch}}\right|.
\end{aligned}
    \label{eq:app-grover-boundary-coupling}
\end{equation}
Choosing the phase of $\ket{\psi_{\rm ch}}$ appropriately gives
\begin{equation}
    H_{\rm eff}^{\rm Grover}\ket{\psi_{\rm ch}}
    =
    E_{\rm ch}\ket{\psi_{\rm ch}}
    +\gamma\ket{\psi_{\rm M}}.
\end{equation}

These identities are exact for any marked set of size $M$.

To account for the remaining feasible states, project the Hamiltonian onto the subspace orthogonal to $\ket{\psi_{\rm ch}}$ within the feasible sector. Let $\ket{\widetilde\psi_{\rm M}}$ be a normalized eigenstate of this projected Hamiltonian, with energy $E_{\rm M}$, representing the dressed marked branch.

We assume an isolated resonance $E_{\rm ch}(\theta_c)=E_{\rm M}(\theta_c)$, with all other eigenvalues of the projected Hamiltonian separated from the crossing by at least $\Delta_{\rm sep}$, where $\gamma/\Delta_{\rm sep}\to0$. Eliminating these nonresonant states gives corrections of order $\gamma^2/\Delta_{\rm sep}$. After choosing the relative phase, the leading Hamiltonian in the basis $\{\ket{\psi_{\rm ch}},\ket{\widetilde\psi_{\rm M}}\}$ is
\begin{equation}
    H_2=
    \begin{pmatrix}
        E_{\rm ch}
        &
        \gamma
        \left|\braket{\psi_{\rm M}|\widetilde\psi_{\rm M}}\right|
        \\[4pt]
        \gamma
        \left|\braket{\psi_{\rm M}|\widetilde\psi_{\rm M}}\right|
        &
        E_{\rm M}
    \end{pmatrix}.
    \label{eq:app-grover-H2}
\end{equation}
At resonance, the diagonal energies coincide, and the splitting is twice the effective coupling:
\begin{equation}
\begin{aligned}
    \Delta_{\rm cross}
    &=
    2\gamma
    \left|\braket{\psi_{\rm M}|\widetilde\psi_{\rm M}}\right|
    \\
    &\quad
    +O\!\left(\frac{\gamma^2}{\Delta_{\rm sep}}\right).
\end{aligned}
    \label{eq:app-grover-crossing-gap}
\end{equation}

The Grover scaling follows under the additional conditions
\begin{equation}
\begin{aligned}
    n\cos^2(\theta_c/2) &= \Theta(1),\\
    \left|
    \braket{n-1|\psi_{\rm ch}}
    \braket{\psi_{\rm M}|\widetilde\psi_{\rm M}}
    \right| &= \Theta(1).
\end{aligned}
    \label{eq:app-grover-scaling-assumptions}
\end{equation}
We assume that the resonance occurs at $\cos^2(\theta_c/2)=\Theta(n^{-1})$. Since $\theta_c$ approaches $\pi$, this implies $\sin\theta_c=\Theta(n^{-1/2})$. The boundary coupling therefore scales as $\gamma=\Theta(\omega\sqrt{M/N})$, provided that $|\braket{n-1|\psi_{\rm ch}}|=\Theta(1)$. The first condition places the resonance near $\theta=\pi$, where the chain's on-site energy spacing is of order $\omega$. Its hopping near the final layer is of order $\omega\sqrt n\sin\theta$; balancing these scales motivates $\sin\theta_c=\Theta(n^{-1/2})$. The second condition requires finite weight of the chain state at its boundary and finite overlap of the dressed marked branch with the uniform marked state. These overlaps depend on the eigenstates and are not fixed by the layer dimensions alone.

Substituting Eq.~\eqref{eq:app-grover-boundary-coupling} yields
\begin{equation}
    \Delta_{\rm cross}
    =
    \Theta\!\left(\omega\sqrt{\frac{M}{N}}\right).
\end{equation}
If the coupling varies negligibly across the resonance and this crossing is the dominant bottleneck, the relevant minimum gap has the same scaling.

\subsection{Local adiabatic schedule}

Within the narrow-crossing approximation, we treat the coupling and diabatic basis as constant. After subtracting the mean energy, the two-level Hamiltonian takes the form
\begin{equation}
    H_2(\theta)
    =
    \frac12
    \begin{pmatrix}
        \delta(\theta) & \Delta_{\rm cross}\\
        \Delta_{\rm cross} & -\delta(\theta)
    \end{pmatrix},
\end{equation}
where $\delta(\theta)=E_{\rm ch}(\theta)-E_{\rm M}(\theta)$. Its instantaneous gap is
\begin{equation}
    \Delta_2(\theta)
    =
    \sqrt{\delta(\theta)^2+\Delta_{\rm cross}^2}.
    \label{eq:app-grover-two-level-gap}
\end{equation}
Writing its instantaneous lower and upper eigenstates as $\ket{E_0}$ and $\ket{E_1}$, respectively, gives
\begin{equation}
    \frac{|\bra{E_1}\dot H_2\ket{E_0}|}{\Delta_2^2}
    =
    \frac{\Delta_{\rm cross}|\dot\delta|}{2\Delta_2^3}.
\end{equation}
The local adiabatic schedule~\cite{RolandCerf2002}
\begin{equation}
    |\dot\delta|=\epsilon\Delta_2^2,
    \qquad
    |\dot\theta|
    =
    \frac{\epsilon\Delta_2^2}{|\partial_\theta\delta|}
    \label{eq:app-grover-local-schedule}
\end{equation}
bounds this ratio by $\epsilon/2$. Here $\epsilon\ll1$ is independent of $n$, and the detuning is assumed to vary monotonically with nonzero slope through the crossing.

Integrating between detunings $\delta_-<0<\delta_+$ gives
\begin{align}
    T_{\rm cross}
    &=
    \frac1\epsilon
    \int_{\delta_-}^{\delta_+}
    \frac{d\delta}{\delta^2+\Delta_{\rm cross}^2}
    \nonumber\\
    &\leq
    \frac{\pi}{\epsilon\Delta_{\rm cross}}
    =
    O\!\left(
        \frac{1}{\epsilon\omega}\sqrt{\frac{N}{M}}
    \right).
    \label{eq:app-grover-runtime}
\end{align}
Thus the two-level crossing reproduces the Grover runtime scaling at fixed adiabatic accuracy. The full protocol retains this scaling up to polynomial factors in $n$ provided that evolution outside the crossing adds at most polynomial overhead and that nonresonant transitions and the velocity-dependent gauge term remain controlled.

\section{\texorpdfstring{Constant gap in the permutation-invariant sector of complete $3$-XORSAT}{Constant gap in the permutation-invariant sector of complete 3-XORSAT}}
\label{app:3xor-gap}

We consider a satisfiable complete $3$-XORSAT instance,
\begin{equation}
    x_i\oplus x_j\oplus x_k=b_{ijk},
    \qquad
    1\leq i<j<k\leq n,
    \label{eq:app-3xor-constraint}
\end{equation}
with $n\geq4$. Throughout this appendix we consider the frozen $g=0$ secondary Hamiltonian, with $\dot\phi=-\omega$ and $\dot\theta=0$, and denote its gap in the symmetry sector accessible from the root state $\ket{u}^{\otimes n}$ by $\Delta_A(n,\theta)$. We show that there exists a constant $c_*>0$, independent of $n$ and $\theta$, such that
\begin{equation}
    \Delta_A(n,\theta)\geq c_*\omega,
    \qquad
    0\leq\theta\leq\pi.
    \label{eq:app-3xor-gap-claim}
\end{equation}

Let $z=(z_1,\ldots,z_n)$ be a satisfying assignment. The relabeling $x_i\mapsto x_i\oplus z_i$ maps the instance to
\begin{equation}
    x_i\oplus x_j\oplus x_k=0,
    \qquad
    1\leq i<j<k\leq n.
    \label{eq:app-3xor-homogeneous}
\end{equation}
In the qutrit Hilbert space, this transformation exchanges $\ket0_i\leftrightarrow\ket1_i$ on sites with $z_i=1$, while leaving $\ket u_i$ invariant. Since the two assignment branches enter the $g=0$ gauge matrix with identical amplitudes, this gives a unitary equivalence of the projected gauge Hamiltonian. The root state is unchanged, so the accessible symmetry sector is mapped unitarily to the permutation-invariant sector of the homogeneous instance. It therefore suffices to prove Eq.~\eqref{eq:app-3xor-gap-claim} for the homogeneous instance.

\subsection{Permutation symmetry and feasible orbits}
\label{app:3xor-symmetry}

The homogeneous complete instance is invariant under arbitrary permutations of the $n$ variables. Denoting the corresponding unitary by $U_\pi$, one has $[U_\pi,P_{\mathcal G}]=0$ and $[U_\pi,A^{(g=0)}]=0$. Hence the permutation-invariant sector is preserved by the $g=0$ gauge dynamics. Since the root state $\ket u^{\otimes n}$ is permutation invariant, its evolution remains in this sector, including when $\dot\theta\neq0$.

For a feasible configuration $\mu$, let $n_0(\mu)=\#\{i:\mu_i=0\}$ and $n_1(\mu)=\#\{i:\mu_i=1\}$, so that $D(\mu)=n_0(\mu)+n_1(\mu)$. Configurations with the same pair $(n_0,n_1)$ form a permutation orbit of size
\begin{equation}
    d_{n_0,n_1}
    =
    \frac{n!}{n_0!\,n_1!\,(n-n_0-n_1)!}.
    \label{eq:app-3xor-orbit-size}
\end{equation}
The corresponding normalized symmetric state is
\begin{equation}
    \ket{n_0,n_1}
    =
    \frac{1}{\sqrt{d_{n_0,n_1}}}
    \sum_{\substack{
        \mu\in\mathcal G\\
        n_0(\mu)=n_0,\;n_1(\mu)=n_1
    }}
    \ket\mu.
    \label{eq:app-3xor-orbit-state}
\end{equation}

For $D<3$, no clause is fully assigned and every partial assignment is feasible. At $D=3$, the assigned triple must contain an even number of ones. The allowed orbit types are therefore
\begin{align}
    D=0 &: \quad (n_0,n_1)=(0,0), \nonumber\\
    D=1 &: \quad (n_0,n_1)=(1,0),(0,1), \nonumber\\
    D=2 &: \quad (n_0,n_1)=(2,0),(1,1),(0,2), \nonumber\\
    D=3 &: \quad (n_0,n_1)=(3,0),(1,2).
    \label{eq:app-3xor-shallow-orbits}
\end{align}

For $D\geq4$, choose any four assigned variables $p,q,i,j$. Feasibility requires $x_p\oplus x_i\oplus x_j=0$ and $x_q\oplus x_i\oplus x_j=0$, and hence $x_p=x_q$. Since $p$ and $q$ are arbitrary, all assigned variables have the same value. Substitution into any assigned triple gives $x\oplus x\oplus x=x=0$. Thus every feasible configuration beyond the first three layers belongs to the all-zero branch,
\begin{equation}
    (n_0,n_1)=(D,0),
    \qquad
    D\geq4.
    \label{eq:app-3xor-deep-orbits}
\end{equation}
The complete set of symmetric feasible orbits is consequently
\begin{equation}
\begin{aligned}
    \mathcal S_n
    &=
    \{(D,0):0\leq D\leq n\}\\
    &\quad\cup
    \{(0,1),(1,1),(0,2),(1,2)\},
\end{aligned}
    \label{eq:app-3xor-orbit-set}
\end{equation}
so that $\dim\mathcal G_{\rm sym}=n+5$.

The chain states are
\begin{equation}
    \ket D\equiv\ket{D,0},
    \qquad
    D=0,\ldots,n.
    \label{eq:app-3xor-backbone}
\end{equation}
They connect the root $\ket0=\ket u^{\otimes n}$ to the unique complete solution $\ket n=\ket0^{\otimes n}$. The remaining four symmetric states are confined to $D\leq3$.

\subsection{Collective-spin chain and shallow orbits}
\label{app:3xor-collective-spin}

After dropping the scalar shift $-n\omega\cos^2(\theta/2)$ and applying $\ket\mu\mapsto(-1)^{D(\mu)}\ket\mu$, the dimensionless Hamiltonian restricted to the chain is
\begingroup\small
\begin{align}
    h_{\rm ch}
    &\equiv \frac{H_{\rm ch}}{\omega}
    =
    \cos\theta\sum_{D=0}^{n}D\ket D\bra D
    \nonumber\\
    &\quad-
    \frac{\sin\theta}{2\sqrt2}
    \sum_{D=0}^{n-1}
    \sqrt{(D+1)(n-D)}
    \left(\ket{D+1}\bra D+\mathrm{H.c.}\right).
    \label{eq:app-3xor-backbone-H}
\end{align}
\endgroup
We identify $\ket D$ with the Dicke state $\ket{J=n/2,m=D-n/2}$. The spin operators then satisfy
\begin{gather}
J_z\ket D=(D-n/2)\ket D,
\nonumber\\
\bra{D+1}J_x\ket D=\frac{\sqrt{(D+1)(n-D)}}{2}.
\nonumber
\end{gather}
Substitution gives
\begin{equation}
    h_{\rm ch}
    =
    \frac n2\cos\theta
    +\cos\theta\,J_z
    -\frac{\sin\theta}{\sqrt2}J_x.
    \label{eq:app-3xor-large-spin-H}
\end{equation}
Thus the chain is exactly a single collective spin $J=n/2$ in a uniform effective field. With
\begin{equation}
    \delta(\theta)
    =
    \sqrt{\cos^2\theta+\frac12\sin^2\theta},
    \label{eq:app-3xor-delta}
\end{equation}
a rotation about the $y$ axis gives $h_{\rm ch}=n\cos\theta/2+\delta(\theta)J_{z'}$. Hence the chain level spacing is
\begin{equation}
    \Delta_{\rm ch}(\theta)
    =
    \omega\delta(\theta)
    \geq\frac{\omega}{\sqrt2}.
    \label{eq:app-3xor-backbone-gap}
\end{equation}

We now restore the four shallow orbit states. Define
\begin{equation}
    v=\cos\theta,
    \qquad
    \tau=\frac{\sin\theta}{2\sqrt2}.
\end{equation}
The full symmetric Hamiltonian takes the block form
\begin{equation}
    h_n\equiv\frac{H_{\rm sym}}{\omega}
    =
    \begin{pmatrix}
        h_{\rm ch} & V\\
        V^\dagger & h_{\rm s}
    \end{pmatrix},
    \label{eq:app-3xor-symmetric-block}
\end{equation}
where
\begin{equation}
    V
    =
    -\tau\left(
        \sqrt n\,\ket0\bra{0,1}
        +
        \sqrt{n-1}\,\ket1\bra{1,1}
    \right).
    \label{eq:app-3xor-backbone-defect-coupling}
\end{equation}
For compactness, let $\alpha=\tau\sqrt{n-1}$ and $\beta=\tau\sqrt{n-2}$. In the ordered basis $\{\ket{0,1},\ket{1,1},\ket{0,2},\ket{1,2}\}$,
\begin{equation}
    \setlength{\arraycolsep}{3.5pt}
    h_{\rm s}
    =
    \begin{pmatrix}
        v & -\alpha & -\sqrt2\,\alpha & 0\\
        -\alpha & 2v & 0 & -\sqrt2\,\beta\\
        -\sqrt2\,\alpha & 0 & 2v & -\beta\\
        0 & -\sqrt2\,\beta & -\beta & 3v
    \end{pmatrix}.
    \label{eq:app-3xor-defect-H}
\end{equation}
Thus all deviations from the collective-spin chain are confined to the first three assignment layers.

\subsection{Uniform gap in the permutation-invariant sector}
\label{app:3xor-uniform-gap}

We prove that the gap of $h_n$ has a positive lower bound independent of $n\geq4$ and $\theta\in[0,\pi]$. The chain eigenvalues are $-L+j\delta$, $j=0,\ldots,n$, where
\begin{equation}
    \delta=\sqrt{v^2+4\tau^2}\geq\frac1{\sqrt2},
    \qquad
    L=\frac n2(\delta-v).
\end{equation}

First consider $L\to\infty$. Let $P$ project onto the two lowest eigenvectors of $h_{\rm ch}$, and set $q=L/(n\delta)$. Their amplitudes at the two sites coupled to $h_{\rm s}$ form the matrix
\begin{equation}
    R=(1-q)^{(n-2)/2}
    \begin{pmatrix}
        1-q & -\sqrt{nq(1-q)}\\
        \sqrt{nq(1-q)} & 1-nq
    \end{pmatrix}.
    \label{eq:3xor-boundary-matrix}
\end{equation}
This follows by rotating the two lowest Dicke states. Since $(1-q)^n\leq e^{-nq}$ and $nq=L/\delta\geq L$, one has $\|R\|\leq C(1+L)e^{-L/2}$. As $V$ acts only at these sites and $\|V\|=\tau\sqrt n\leq\sqrt L$,
\begin{equation}
    \|PV\|
    \leq C\sqrt L(1+L)e^{-L/2}
    =o(1).
    \label{eq:3xor-decoupling-matrix}
\end{equation}

It remains to exclude low-energy states formed by the complementary chain modes and the four additional states. Write $K$ for the compression of $h_n$ to this complementary space. Its chain block is bounded below by $-L+2\delta$, while the row-sum bound on the four-dimensional block gives
\begin{equation}
    \lambda_{\min}(h_{\rm s})
    \geq-3-(1+\sqrt2)\tau\sqrt n.
\end{equation}
The quadratic-form comparison yields
\begin{align}
    \lambda_{\min}(K)+L
    &\geq
    \lambda_{\min}
    \begin{pmatrix}
        2\delta & -\tau\sqrt n\\
        -\tau\sqrt n &
        L-3-(1+\sqrt2)\tau\sqrt n
    \end{pmatrix}
    \nonumber\\
    &\geq
    2\delta-1-O(L^{-1/2}),
    \label{eq:3xor-complement-matrix}
\end{align}
where the last inequality follows from $n\tau^2=L(\delta+v)/2\leq L$ and
\begin{equation}
    \lambda_{\min}
    \begin{pmatrix}
        a & -b\\
        -b & d
    \end{pmatrix}
    \geq a-\frac{b^2}{d-a},
    \qquad d>a.
\end{equation}
Removing $PV$ leaves the two chain energies $-L,-L+\delta$ and the complementary spectrum bounded below by Eq.~\eqref{eq:3xor-complement-matrix}. Weyl's inequality therefore gives
\begin{equation}
    \liminf_{L\to\infty}\frac{\Delta_A}{\omega}
    \geq\sqrt2-1>0.
    \label{eq:3xor-large-L-gap}
\end{equation}

Next let $n\to\infty$ with bounded $L$. Then $v\to1$ and $\sqrt n\,\tau$ is bounded. Along a subsequence with $\sqrt n\,\tau\to\kappa$, the matrix converges entrywise at fixed depth to
\begin{equation}
    h_\infty(\kappa)
    =
    \begin{pmatrix}
        h_{{\rm ch},\infty} & V_\infty\\
        V_\infty^\dagger & h_{{\rm s},\infty}
    \end{pmatrix}.
    \label{eq:3xor-limit-matrix}
\end{equation}
Here $h_{{\rm ch},\infty}$ is tridiagonal, with
\begin{equation}
\begin{aligned}
    (h_{{\rm ch},\infty})_{D,D}&=D,\\
    (h_{{\rm ch},\infty})_{D,D+1}
    &=-\kappa\sqrt{D+1}.
\end{aligned}
\end{equation}
The remaining blocks follow from the matrices above by taking $v\to1$ and $\tau\sqrt n,\alpha,\beta\to\kappa$.

To justify the low-energy limit, let $\widehat D$ denote the diagonal depth operator, and write $\kappa_n=\sqrt n\,\tau$. The hopping block between depths $D$ and $D+1$ has norm at most
\begin{equation}
    \tau\sqrt{2(n-D)(D+1)}
    \leq
    \sqrt2\,\kappa_n\sqrt{D+1}.
\end{equation}
This follows from the maximal forward and backward degrees $2(n-D)$ and $D+1$, respectively. For any normalized vector $\psi$, Cauchy--Schwarz therefore gives
\begin{equation}
\begin{aligned}
    \langle\psi,h_n\psi\rangle
    &\geq
    v\langle\widehat D\rangle_\psi
    -2\sqrt2\,\kappa_n
    \sqrt{\langle\widehat D\rangle_\psi}\\
    &\geq
    \frac12\langle\widehat D\rangle_\psi-C,
\end{aligned}
    \label{eq:3xor-depth-confinement}
\end{equation}
with $C$ independent of $n$ along the sequence.

The min--max principle applied to the bounded principal matrix
\begin{equation}
    \begin{pmatrix}
        0 & -\kappa_n\\
        -\kappa_n & v
    \end{pmatrix}
\end{equation}
gives a uniform upper bound on the first excited energy $\lambda_1(h_n)$. Thus both normalized low-energy eigenvectors satisfy
\begin{equation}
    \langle\widehat D\rangle_\psi\leq C',
    \qquad
    \sum_{D>\ell}\|\psi_D\|^2
    \leq\frac{C'}{\ell+1},
\end{equation}
where $\psi_D$ collects the amplitudes at depth $D$. Embedding the finite matrices into $\ell^2(\mathbb N_0)\oplus\mathbb C^4$, this bound makes the low-energy eigenvectors precompact in norm.

The off-diagonal part of $h_\infty$ grows only as $\sqrt D$, so it is infinitesimally $\widehat D$-bounded, and its product with $(\widehat D+1)^{-1}$ is compact. Thus $h_\infty$ is self-adjoint with compact resolvent and has finite-support vectors as a core.

Suppose the gap vanished along the subsequence. The two orthogonal low-energy eigenvectors would have strongly convergent subsequences, with orthogonal limits satisfying the eigenvalue equation for $h_\infty$ at a common energy $E$. For every finite-support vector $\varphi$, the variational principle and fixed-depth matrix convergence imply
\begin{equation}
    E\leq
    \frac{\langle\varphi,h_\infty\varphi\rangle}
    {\|\varphi\|^2}.
\end{equation}
Since $E$ is an eigenvalue, this yields $E=\inf\sigma(h_\infty)$.

This is impossible. For $\kappa>0$, the limiting matrix is connected with strictly negative hopping. Its positivity-improving semigroup and compact resolvent imply a simple ground eigenvalue~\cite{ReedSimon1978}. For $\kappa=0$, the depth-zero root is its unique ground state.

For fixed $n$, the same irreducibility argument applies for $0<\theta<\pi$; at $\theta=0,\pi$, the gaps equal $\omega$. Continuity excludes gap closing at fixed $n$. Any gap-closing sequence must therefore have $n\to\infty$ and, after taking a subsequence, either $L\to\infty$ or bounded $L$. Both possibilities have been excluded. Hence
\begin{equation}
    \inf_{\substack{n\geq4\\0\leq\theta\leq\pi}}
    \frac{\Delta_A(n,\theta)}{\omega}>0.
\end{equation}

\section{Classical computational budgets}
\label{app:classical-budget}

We examine the dependence of classical performance on the clause-inspection budget. For these $3$-SAT benchmarks, time is measured in units of $\omega^{-1}$, with $\omega=1$ in the numerical calculations. All methods use the same 30 SAT instances per size and ensemble, with clause density $m/n=6$ and exact optima determined by enumeration. Quantum evolution uses $T=n^2$, followed by measurement and uniform completion of any unassigned variables. All reported scores refer to complete assignments.

The classical budget $B$ counts clause inspections, including initialization and rejected proposals, rather than complete formula evaluations or elapsed time. Evaluating an initial complete assignment costs $m=6n$ inspections. A single-variable proposal inspects the clauses containing that variable, costing 18 inspections in the regular ensemble. Consequently, in the regular ensemble, Greedy, Metropolis, and simulated annealing (SA) can attempt at most
\begin{equation}
K_{\mathrm{proposal}}
\leq
\left\lfloor\frac{B-6n}{18}\right\rfloor
\end{equation}
local updates after initialization. At the main-text budget $B=n^2$, no local update is possible for $n=7,8$, and at most seven are possible for $n=15$. Thus, this budget substantially restricts classical search. Random returns the best of $\lfloor B/m\rfloor$ complete assignments. Random partial assignment reserves $m$ inspections for final scoring, truncates construction at its budget, and uniformly completes all remaining unassigned variables.

\begin{figure}[!htbp]
\centering
\includegraphics[width=\columnwidth]{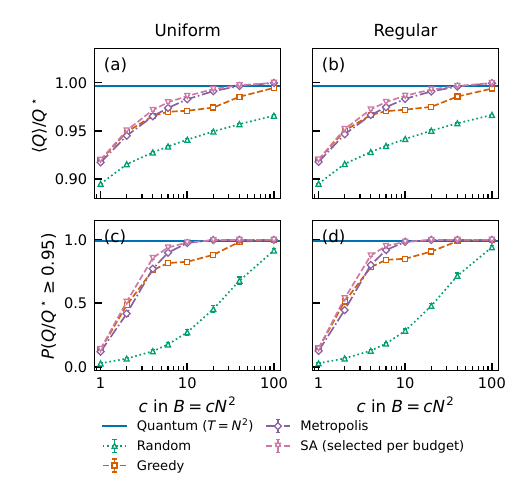}
\caption{
Classical budget dependence for SAT instances at $n=15$ and $m/n=6$. Columns correspond to uniform and regular ensembles. (a),(b) Mean approximation ratio $\langle Q\rangle/Q^\star$. (c),(d) Probability $P(Q/Q^\star\geq0.95)$. The horizontal axis gives the classical clause-inspection cap $B$; the dotted vertical line marks the main-text budget $B=n^2=225$. Blue horizontal lines and shaded bands show the fixed quantum means and one standard error at $T=225$. SA schedules are selected separately at each budget.
}
\label{fig:app-budget-sat}
\end{figure}

To assess this restriction, we fix $n=15$ and vary $B=cn^2$ with $c\in\{1,2,4,6,10,20,40,100\}$. Figure~\ref{fig:app-budget-sat} reports the actual clause-inspection cap $B=225$--$22\,500$ on a logarithmic horizontal axis. The quantum reference remains fixed at $T=n^2=225$ throughout. The classical methods are Random, Greedy, Metropolis with $\beta=1$, and SA. SA schedules are selected separately at each budget from 17 candidates using 144 disjoint training instances; tuning costs are additional to the reported test-run budget. This differs from the fixed $0.1\to0.01$ schedule used in the main-text comparison. Stochastic results use 1,024 runs per test instance, while Random is evaluated exactly. Error bars denote one standard error across instances.

Classical performance improves substantially with increasing budget. At $B=22\,500$, the SA mean approximation ratio exceeds the fixed quantum mean in both ensembles. UNSAT instances exhibit the same qualitative budget dependence. The main-text comparison therefore characterizes finite-size performance under the specified short classical budget, rather than superiority over classical algorithms given sufficient search time. Since quantum evolution time and classical clause-inspection counts measure distinct resources, these results do not establish an equal-cost computational advantage.

%
%
%

\end{document}